\documentclass[%
 reprint,
superscriptaddress,
nofootinbib,
 amsmath,amssymb,
 aps,
floatfix,
]{revtex4-2}

\usepackage{graphicx}
\usepackage{dcolumn}
\usepackage{bm}
\usepackage{float}
\usepackage{subcaption}
\usepackage{xurl}

\usepackage[sort&compress]{natbib}
\usepackage{multirow}

\usepackage[usenames,dvipsnames]{xcolor}
\definecolor{xlinkcolor}{cmyk}{1,1,0,0}

\usepackage[
 colorlinks=true,    
 linkcolor=xlinkcolor,     
 citecolor=xlinkcolor,     
 filecolor=xlinkcolor,  
 urlcolor=xlinkcolor,      
 final=true
]{hyperref}

\usepackage{pdfpages}
\makeatletter
\AtBeginDocument{\let\LS@rot\@undefined}
\makeatother

\begin{document}

\preprint{APS/123-QED}

\title{Transforming U.S. Particle Physics Education: A Snowmass 2021 Study}

\author{O.~Bitter} \affiliation{Fermi National Accelerator Laboratory, Neutrino Division, Batavia, IL 60510, USA}\affiliation{The University of Chicago, Physical Science Division, Chicago, IL 60637, USA}  
\author{E.V.~Hansen} \affiliation{University of California Berkeley, Department of Physics, Berkeley, CA 94720, USA}  
\author{S.~Kravitz} \affiliation{Lawrence Berkeley National Laboratory, 1 Cyclotron Rd., Berkeley, CA 94720, USA}  
\author{V.~Velan} \affiliation{University of California Berkeley, Department of Physics, Berkeley, CA 94720, USA}  
\author{Y.~You} \affiliation{University of Florida, Department of Physics, Gainesville, FL 32601, USA}


\date{\today}

\begin{abstract}

 \textbf{ \hspace{5cm} Abstract}

The pursuit of knowledge in particle physics requires constant learning. As new tools become available, new theories are developed, and physicists search for new answers with ever-evolving methods. However, it is the case that formal educational systems serve as the primary training grounds for particle physicists. Graduate school (and undergraduate school to a lesser extent) is where researchers learn most of the technical skills required for research, develop scientific problem-solving abilities, learn how to establish themselves in their field, and begin developing their career. It is unfortunate, then, that the skills gained by physicists during their formal education are often mismatched with the skills actually required for a successful career in physics. We performed a survey of the U.S.~particle physics community to determine the missing elements of graduate and undergraduate education and to gauge how to bridge these gaps. In this contributed paper, part of the 2021-22 Snowmass Community Planning Exercise, we report the results of this survey. We also recommend several specific community actions to improve the quality of particle physics education; the ``community'' here refers to physics departments, national labs, professional societies, funding agencies, and individual physicists.

\end{abstract}

\pacs{Valid PACS appear here}

\maketitle

\renewcommand{\thefootnote}{\arabic{footnote}}




\section*{Executive Summary}
\label{sec:ExecutiveSummary}

\begin{enumerate}
\item \textbf{Finding:} Survey respondents are mostly satisfied with their graduate education (Section~\ref{sec:gradsatisfaction}) and feel relatively well-prepared for careers in academia or in laboratories (Section~\ref{sec:careerprep}). However, perceived preparation is moderate for careers in industry and worse for K-12 education, including among those who intend on these non-academic careers (Section~\ref{sec:careerprep}). \\

\textbf{Recommendation:} Graduate programs in particle physics should normalize training for industry positions via encouragement of industry partnerships (such as summer research internships) and formal development of skills in-demand beyond academia (such as computer programming, team/project management, and effective communication).

\item \textbf{Finding:} Survey respondents were most likely to be interested in academic careers when starting graduate school, though a sizeable number have since changed career intention to laboratory, industry, or other careers (Section~\ref{sec:gradcareers}). Furthermore, a sizeable number of survey respondents report changing intention from HEP theory to HEP experiment (Section~\ref{sec:subfieldprefs}). Lack of positions and/or funding were often cited as the reason for this. 
\\

\textbf{Recommendation:} Universities should provide undergraduate students with a more complete picture of what particle physicists do beyond classroom discussion of physics theory, such as increased opportunities for learning about research (e.g. seminars). They should also provide a more realistic view of common career paths post-PhD in particle physics, including the breakdown of theory and experimental academic positions as well as the commonality of shifting to a non-academic career (e.g. through job panels). This could help students make a more informed decision about what to study in graduate school and whether such a choice aligns with their goals.

\item \textbf{Finding:} Professional skills, such as technical presentations and scientific writing, are considered important for one's career and are more strongly-correlated with reported career preparation than technical skills such as computer programming (Section~\ref{sec:profskills}, \ref{sec:preparedness}). However, they are very frequently gained through self-teaching rather than through any sort of training, including peer learning or mentoring; this self-taught mode is largely disfavored by respondents, while alternative modes such as university courses are more highly-rated. There is some evidence that physicists are being more formally trained in these skills over time, though this effect is small.\\

\textbf{Recommendation:} Graduate programs in particle physics should support more formal modes of training for those skills where self-teaching is inadequate, in favor of the preferred mode as indicated in Table~\ref{tab:pvalues}. As many of these professional skills are equally useful throughout physics disciplines or even other scientific and non-scientific fields, this could take many forms such as university-wide workshops, one-on-one coaching sessions, or shared online resources. It is critical for advisors or program coordinators to make their students aware of such resources and to actively encourage their use as part of their graduate training (and not a ``free time'' activity).

\item \textbf{Finding:} Despite receiving high career importance ratings, computer programming and statistics remain somewhat likely to be learned through self-teaching or peer learning (Sec~\ref{sec:techskills}). Among theorists, theoretical mathematical skills were also somewhat commonly received through these methods (Sec~\ref{sec:mathskills}). There is evidence that these methods are poorly rated, especially for statistics, while learning through a university course is rated well. \\

\textbf{Recommendation:} Physics departments should consider making a course in statistics a standard part of the undergraduate physics curriculum, as well as providing avenues for formally training graduate students in statistics for particle physics. More formal training opportunities should also be made available for advanced theoretical skills, including opportunities outside of the classroom such as virtual workshops which are free to attend.

\item \textbf{Finding:} Only a small subset of survey respondents were undergraduates. We were unable to provide deeper analyses of undergraduate course preparation and career paths due to lack of survey participants. \\

\textbf{Recommendation:} Funding agencies and professional societies should develop connections and networking opportunities - including student oriented conferences - to help undergraduate students remain connected in the HEPA community. Develop a mass communication system to reach undergraduates in the future. Community members should actively plan to perform this survey again in the future for undergraduate data once the communication barriers listed in other findings are rectified.

\item \textbf{Finding:} None of the survey respondents who self-identified as undergraduates reported research experience in an REU or DOE sponsored lab internship. Most undergraduates participating in this survey came from R1 research institutions. \\

\textbf{Recommendation:} funding agencies and national labs should compile a list of contact information for recent undergrad students in internship and REU programs. Create a central mass communication system for physics departments across a variety of college or university types. 

\end{enumerate}


\section{Introduction \& Motivation}
\label{sec:Introduction}

By the most recent metrics \cite{mulveyTrendsPhysicsPhDs2021}, the U.S. conferred just under 2000 physics PhDs in 2019, or 3\% of all PhDs awarded in the U.S.. Of those, roughly 15\% were in `particles and fields', 15\% were in `astronomy, astrophysics, and cosmology', and 7\% were in `nuclear physics'. These graduate students make up a substantial portion of the high energy physics and astrophysics (HEPA) community. 

In order to understand the current particle physics educational climate and accurately conclude how to best improve it, we have conducted a survey of the particle physics community aimed at addressing the status of said curriculum at the undergraduate and graduate level. We have considered the extent to which factors such as a scholar's institution type (public, private, etc.), training and skills (academic, professional, and technical), and satisfaction are linked to their preparation for a career in particle physics. We have also considered how students are being prepared for careers outside of traditional academia or lab-based professions. We did this to inform physics programs of important missing elements in particle physics preparation for their students to be successful---whether it be in an academic, private, or industry-related job sector. This project was completed as part of the 2021 Snowmass Community Planning Exercise (``Snowmass''), a decadal process in which the U.S.~particle physics community builds a 20-year vision of the future of particle physics. Details can be found at \url{https://www.snowmass21.org/}. This paper and the accompanying survey are housed within the Community Engagement Frontier of Snowmass, in Topical Group 4: Physics Education.

\subsection{Context}

In a survey of initial post-degree employment of graduate physics students conducted by the AIP Statistical Research Center \cite{EmploymentCareersPhysics2020}, 49\% of  respondents reported that they had entered the academic sector, 32\% had entered the private sector, and 15\% had entered the government sector. Without a detailed breakdown of these statistics by sub-field, we must assume that HEPA aligns with these data. Those physicists who moved out of physics or astronomy for their initial job primarily moved to engineering (19\%), computer software (15\%), data science (13\%), or other STEM (8\%). 

When the same survey was presented to undergraduate researchers, only 48\% of respondents reported that they had entered graduate study, and only 65\% of those new graduate students had remained in physics (other topics were dominated by engineering, but also included computer science, mathematics, and education). Those who did not enter graduate school after their Bachelor's reported employment in engineering (35\%), computer software (24\%), non-STEM (22\%), and STEM (15\%) jobs.

Physics departments, even if they do not know this, are training students who will exit academia and enter industry. Are our topics of learning reflecting this pathway? Are we training our students to perform in environments that aren't academia, even if these same students were heavily involved in HEPA collaborations during their education? 

We encourage readers to review the data on knowledge and skills used by recently graduated undergraduate \cite{EmploymentCareersPhysics2020} and graduate \cite{mulveyPhysicsDoctoratesSkills} physics students as compiled by the AIP Statistical Research Center. We would like to highlight the following: 
\begin{itemize}
    \item Recent physics PhDs who moved into postdoctoral positions identified several common technical skills used `daily or weekly': \textit{advanced math} ($>$60\%), \textit{advanced physics principles} ($>$75\%), \textit{programming} ($>$75\%). Recent physics PhDs who moved into academic, non-postdoctoral positions used the above skills less frequently. 
    \item Recent physics PhDs who moved into postdoctoral positions identified several common interpersonal skills used `daily or weekly': \textit{work on a team} ($>$75\%), and \textit{technical writing} ($>$50\%). More than 75\% of respondents who moved into academic, non-postdoctoral positions also reported \textit{work[ing] on a team} daily or weekly, but also reported frequent use of skills relating to \textit{speaking publicly} ($>$50\%) and \textit{teaching} ($>$75\%). 
    \item More than 50\% of recent physics PhDs that moved to industry (in physics, engineering, and computer science) reported using \textit{technical writing} `daily or weekly'. These same respondents also reported that they used the following daily or weekly: \textit{solve technical problems} ($>$75\%), \textit{programming} ($>$75\%), \textit{advanced physics principles} ($>$75\%), and \textit{advanced math} ($>$50\%)
    \item Recent physics bachelors that moved to industry identified several common skills as used `daily or weekly'. These skills are probably familiar to those of us in academic careers --- \textit{work on a team} ($>$75\%), \textit{solve technical problems} ($>$75\%), and \textit{technical writing} ($>$50\%) overlap strongly with what we expect academic or laboratory professionals to have proficiency in.
\end{itemize} 

\begin{figure}
    \centering
    \includegraphics[width=\linewidth]{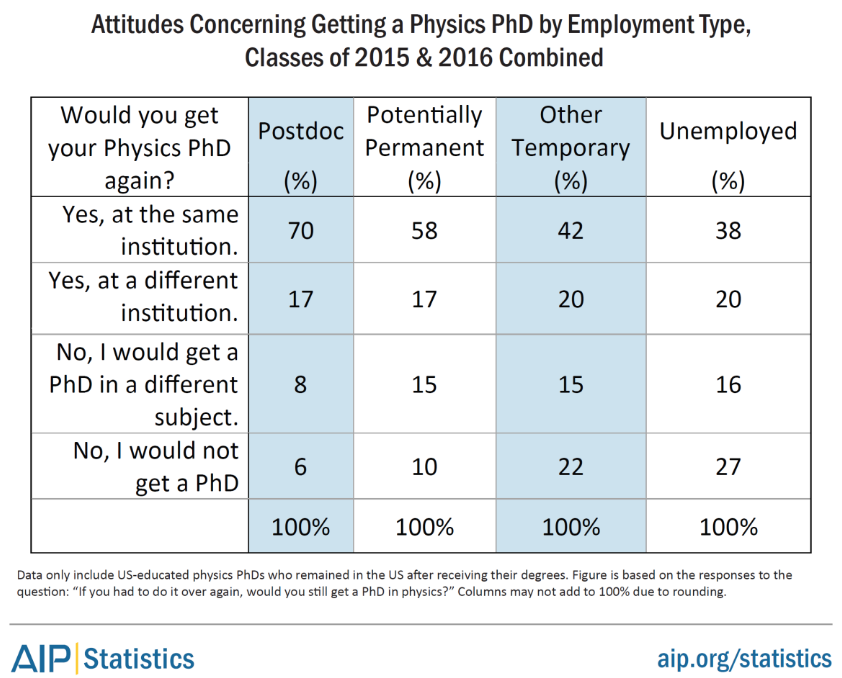}
    \caption{Data from AIP Statistical Research Center from \cite{mulveyPhysicsDoctoratesSkills} on US-educated Physics PhDs from classes of 2015 and 2016, asked \textit{if you had to do it over again, would you still get a PhD in physics?}. This lends credibility to the discussion of subfield and career trajectory changes in sections \ref{sec:gradcareers} and \ref{sec:subfieldprefs}.}
    \label{fig:AIPRepeatPhDQuestion}
\end{figure}

While the survey evaluated for this article did not tie satisfaction in graduate education to any particular variable, it is useful to acknowledge that the AIP Statistical Research Center did ask if recent physics PhDs felt that their advisor was helpful in career planning or advancement. 84\% of those in postdoctoral positions felt that their advisor was `helpful' or `very helpful', compared to 54\% of those in potentially permanent positions. This was attributed to the resources available to university faculty advisors, such as ``networking opportunities and collaboration with colleagues at other universities"\cite{mulveyPhysicsDoctoratesSkills}. This suggests that HEPA advisors should avail themselves of the available resources to mentor students who do not intend to follow an academic trajectory, and not rely on these students to find their own ways. Additionally, the AIP survey also asked if students would repeat their experience in graduate education, which ties in closely with the discussion on switching career paths (see section \ref{sec:subfieldprefs}). We include their results verbatim in Figure~\ref{fig:AIPRepeatPhDQuestion}. 

\subsection{Review of Literature}
Irving and Sayre (2015)\cite{irvingBecomingPhysicistRoles2015} studied the development of a ``physics identity" (the concept of ``being a physicist'' and the student's perspective of their ties to the field), which ties strongly with persistence in physical science \cite{hausslerInterventionStudyEnhance2002, Hazari2013TheSI}. Specifically, they looked at this development through the lens of ``communities of practice", which as defined in \cite{irvingBecomingPhysicistRoles2015} include physics classrooms, academic departments, laboratory divisions, and  HEPA collaborations. The categories tied with a physics identity include ``doing research" (including development of research as a PI), ``mastery of the [research] subject", and ``mastery of the [curriculum] subject''. They note specifically that students who perceive themselves as lacking in these skills perceive themselves as not being ``internal members'' of the community of practice. Later interviews with the same students revealed development towards their physics identity was strongly tied with involvement in research. This suggests that skills developed in the modalities ``in a course organized outside of the classroom'', which may include summer schools and HEPA collaboration workshops, and ``mentoring or peer learning'' (discussed in detail in Section~\ref{sec:gradskills}) should produce stronger ties with physics identity and lead to persistence in physics. 

In a review of the PhD Plus 10 Study \cite{porterPhysicsPhDsTen2019}, a study from the AIP Research Center of Physics PhDs 10 years after their degree, physicists reported that success was attributed not only to ``personal drive'', but also to experience in basic \& advanced physics and mathematics, experience in research, the ability to change employment (including away from academia), and social support in the form of physics collaborations and mentoring relationships. Among the skills listed as critical were technical communication to scientific and non-scientific audiences, teaching, mentoring, programming, and grant writing --- all skills that we discuss later in this article (Sec \ref{sec:profskills}). 

The authors would like to emphasize that one of the most frequent reported barriers to careers post-PhD was ``issues with education experience". Namely, we repeat these quotes verbatim from \cite{porterPhysicsPhDsTen2019}: 
\begin{quote}
    ``Graduate training in physics leaves most students with advanced degrees unprepared for the jobs most will eventually end up with."\\
    ``Lack of understanding of job options coming out of graduate school."\\
    ``Lack of experience in the fields outside of Physics."
\end{quote}
\noindent When considering the development of curriculum within the HEPA community, are we most adequately serving the students that report these kinds of issues? 

The concept that academic institutions are failing to prepare Physics PhDs for industrial positions is not a new problem. In fact, in a report to the American Physical Society by their Economic Concerns Committee in 1971, the following recommendation was made: 
\begin{quote}
    In the next decade a larger share of the jobs for PhD physicists will have to come from outside the educational institutions. To this end programs to increase the involvement of the physics community in industry must be encouraged and stimulated.... Physics departments must reexamine their training programs especially for careers for which few employment opportunites exist. Is it too much to ask that a physicist who gets a PhD in elementary particle physics demonstrate a competence in research in an unrelated physics area?.... We should advise [students] towards a well-grounded preparation in fundamentals, carried through with the broadest of attitudes and the widest of visions. We expect that a physics training based on such a foundation, however narrow may be the thesis topic, will better prepare a [physicist] for a future scientific career than would any alternative training. 
    
    \hfill Grodzins (1971) \cite{grodzinsManpowerCrisisPhysics1971}
\end{quote}

\noindent If it is true that we as the HEPA community must train our undergraduate and graduate students for careers beyond the (commonly thought of as) traditional academic pathway, a) are we doing it enough and b) are we doing it in the right way? 

We also point readers to \cite{Snowmass_Careers_Whitepaper}, another paper within the auspices of Snowmass. In that paper, the Working Group on Career Pipeline and Development analyzed the results of a survey targeted at early-career physicists (a separate survey than the one presented here), focusing on how to support physicists who transition out of HEPA. Factors for this career transition were identified, including available positions, salary, and burnout. The paper also investigated the best strategies to support physicists who make the transition---mentorship from older physicists, communication of available jobs, and permission to engage in non-academic projects and internships. These opportunities should be especially available at key career transition points, i.e. after finishing a degree or other position. As we will discuss in Sections~\ref{sec:careerprep}~and~\ref{sec:preparedness}, training junior scientists during graduate school is another important method to support those who go to industry (even before they might have decided to make the transition). 

Finally, we highlight a recent study in physics education research supporting the development of skills in informal physics programs: namely, that students who participated in the development of K-12 outreach events reported increased development of ``21st century career skills" like communication, network, and design \cite{rethmanImpactInformalPhysics2021}. This supports the hypothesis put forward by this work that such skills, and also the concept of `physics identity', are often developed outside of standardized coursework. The authors would also like to note that comparisons of teaching modalities like self-directed, community-directed, or formal coursework are hard to come by. This is likely due to the breakdown between these sectors when considering what is necessary to succeed in a classroom versus a HEPA collaboration, and which skills each community believes they are responsible for. 


\section{Survey Methodology}
\label{sec:SurveyDetailsAndMethodology}

\begin{table*}
    \begin{tabular}{p{0.2\textwidth} p{0.3\textwidth} p{0.2\textwidth} p{0.2\textwidth}}
                                      &                         & \begin{tabular}[c]{@{}l@{}}Number of Responses\end{tabular} & \begin{tabular}[c]{@{}l@{}}Percent of Responses\end{tabular} \\ \hline\hline\\
    \multirow{8}{*}{Current Position} & Undergraduate Student   &  24   &   6.7     \\
                                      & Graduate Student        &  93   &   26.1    \\
                                      & Postdoctoral Researcher &  56   &   15.7    \\
                                      & Tenure-Track Faculty    &  19   &   5.3     \\
                                      & Tenured Faculty         &  88   &   24.6    \\
                                      & Research Staff$^a$          &  69   &   19.3    \\ 
                                      & Technician/Engineer$^a$     &  8    &   2.3     \\ \\\hline \\
    \multirow{7}{*}{Affiliated Institution(s)} & Public University       &  150  &   42.0    \\
                                      & Private University      &  102  &   28.6    \\
                                      & National Laboratory     &  116  &   32.5    \\
                                      & Liberal Arts College    &  15   &   4.2     \\
                                      & Community College$^b$       &  1    &   0.3     \\
                                      & For-Profit Institution$^b$  &  4    &   1.1     \\
                                      & Other (e.g. CERN)$^b$       &  5    &   1.4     \\ \\\hline
    \end{tabular}
    \caption{Demographic makeup of survey respondents. Categories with fewer than 10 members are either (a) combined with other categories, or (b) excluded from deeper correlation analysis to avoid identification of respondents. Respondents were able to select more than one affiliated institution.} \label{tab:survey_demographics}
\end{table*}

The survey discussed in this work can be found in Appendix \ref{app:full_survey}. It was conducted on the Qualtrics platform\footnote{The survey for this paper was generated using Qualtrics software,  January 2022 Version of Qualtrics. Copyright \copyright~ 2022 Qualtrics. Qualtrics and all other Qualtrics product or service names are registered trademarks or trademarks of Qualtrics, Provo, UT, USA. https://www.qualtrics.com} for 22 days, starting on 14 February 2022. Survey links were sent out to Snowmass email lists and major particle physics collaborations and publicized on the Snowmass Slack workspace. A total of 422 responses were started; 357 responses were finished and included in the results printed here. Two responses were flagged as potential bots by the Qualtrics RECAPTCHA algorithm but were ultimately included after manual review. Due to constraints with local human research ethics protocols\footnote{UC Berkeley CPHS protocol number: 2022-01-14952}, respondents were required to acknowledge that they were currently within the United States, so this survey does not include those who might be affiliated with institutions outside of the U.S.. 

Current position and affiliated institution(s) were the only demographic information collected; these data are summarized in Table \ref{tab:survey_demographics}. Categories with fewer than ten (10) respondents were either combined with other categories (e.g.,~``Technician/Engineer" was combined with ``Research Staff") or excluded from deeper correlations (``Community College", ``For-Profit Institution", and ``Other (e.g. CERN)") to avoid identification of respondents. 

Respondents were required to list their current position in order to process their participation correctly, but otherwise most questions were optional. As such, the number of respondents who answered each question varies; attempts have been made to include the number of respondents displayed in each table and chart.

\subsection{Limits on this analysis}
The survey was primarily directed to the Snowmass community and was therefore advertised on General, Early Career, and Community Engagement spaces. It was also sent out to roughly fifty experiment collaborations in HEPA and six national laboratories with instructions to forward the survey to their respective list-servs. While this does roughly cover the target population, we must refrain from overgeneralizing these results to populations beyond Snowmass and those actively involved in either HEPA experimental collaborations or national laboratories. Specifically, this population is limited in both theory-focused scholars (demonstrated most starkly in Section~\ref{sec:mathskills}) and those who have left academic collaborations for industry focused careers (clearest in Sections~\ref{sec:gradcareers}~and~\ref{sec:subfieldprefs}). 

The number of participants was relatively high compared to other Snowmass surveys (357 completed responses) but is still a low number compared to the total community population. For reference, this is the same order of magnitude as the 488 U.S.-based authors on the ATLAS collaboration alone \cite{USAtlasCollaboration} and is about 7\% of all registered users on the Snowmass2021 Slack workspace\footnote{It is unlikely that all Snowmass2021 Slack workspace users are active in the Snowmass Community Report process, and there are likely scientists involved in the process who are not on Slack, but the population of registered users is certainly a better estimate of the size of the HEPA community.}. 

This survey also did not collect demographic data beyond position and affiliated institution, due to the limits on our human subjects research protocol. It is well known that minoritized populations experience STEM communities differently, both inside and outside of the classroom \cite{goodProblemsPipelineStereotype2007, beasleyWhyTheyLeave2012, lewisFittingOptingOut2016, eddyNumbersReviewGender2016, spencerStereotypeThreatWomen1999, carloneUnderstandingScienceExperiences2007, cooperComingOutClass2016, jonesImpactEngineeringIdentification2013}, which this survey cannot account for. Further study is needed at the graduate level to understand the most impactful methods of learning across identity groups to prepare them for careers within and without the HEPA community. 

There are also a few areas for improvement in question wording. First, a Qualtrics error resulted in the accidental exclusion of an ``Other'' option for career position, which likely excluded non-tenure-track faculty. Second, in the final question of the Undergraduate Education survey, we accidentally listed the number 6 twice for self-reported career preparation. We expect that few respondents would have been confused by this, because the question layout makes the intention obvious. Third, we asked survey respondents to rate the importance of various professional, technical, and mathematical skills in their career, but we only asked respondents who had not received these skills in formal schooling. It would have been more useful to ask all respondents this question.

Finally, the main limit to this survey that we must acknowledge is that this project was designed and analyzed by students and postdoctoral researchers in physics. Our expertise in human behavior is limited, and this skewed both our development of the survey as well as our analysis. We detail a set of `lessons learned' in section \ref{sec:missingquestions}. The authors propose that this survey should not be used as the definitive work in this area, but rather a launching point for further discussion into the modalities of learning in HEPA communities. 


\section{Graduate Education Results}
\label{sec:gradresults}

All respondents who did not select ``Undergraduate Student" in the first question, ``\textit{What position do you currently hold?}", were shown the Graduate Education version of the survey. This means that 333 Graduate Education responses were collected, as shown in Table~\ref{tab:survey_demographics}. In this section, when we refer to `all' participants, unless otherwise indicated, we mean these 333 respondents. 

We collected a variety of ratings for satisfaction with graduate education, preparation for career, and skill modalities. We then compared these to other quantities such as demographic information (position and current institution affiliation) using the Chi-square ($\chi^2$) test of independence, e.g. with null hypothesis $H(0)=$~`response does not depend on position / current institute affiliation'. Unless otherwise noted, $p$-values refer to this test. 

We also quantify the strength of correlation in certain studies (e.g. Section~\ref{sec:preparedness}) using Kendall's Tau \cite{kendallCorrelation}, as it is a suitable metric for ranked categorical data where intervals between categories may not be equal. We note that the numerical value of $\tau$ is smaller than that of other correlation coefficients such as the standard linear correlation coefficient (Pearson's R); to help in the interpretation of these values, we take a range of $0.1<|\tau|\le0.2$ to indicate a weak correlation, $0.2<|\tau|\le0.3$ to be moderate, and $|\tau|>0.3$ to be a strong correlation.

\subsection{Satisfaction with Graduate Education}
\label{sec:gradsatisfaction}

\begin{figure}
    \centering
    \includegraphics[width=3.4in]{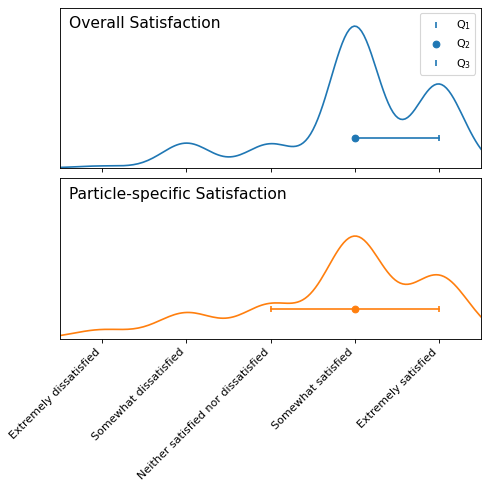}
    \caption{Satisfaction with graduate education for all post-undergrad respondents. Overall satisfaction is shown in blue, particle-specific in orange.  Quartiles are also presented.}
    \label{fig:GradSatisfactionOverall}
\end{figure}

\begin{figure}
    \centering
    \includegraphics[width=\linewidth]{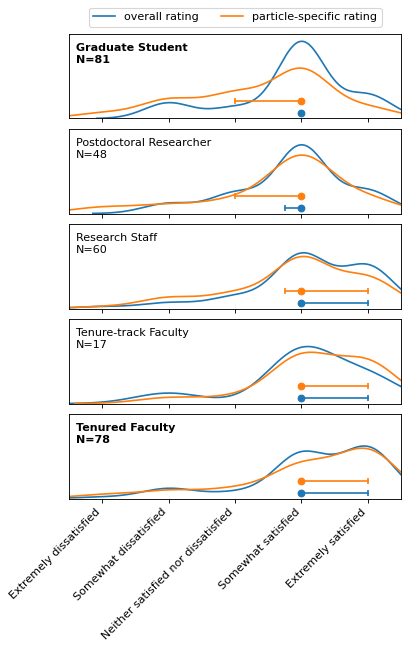}
    \caption{Satisfaction with graduate education as a function of current position. Overall satisfaction is shown in blue, particle-specific in orange.  Quartiles are also presented. The \textbf{bolded} categories are those for which there is a statistically significant relationship between categories and rating ($p < 0.05$ in a $\chi^2$ test of independence).}
    \label{fig:GradSatisfactionByPosition}
\end{figure}

\begin{figure}
    \centering
    \includegraphics[width=\linewidth]{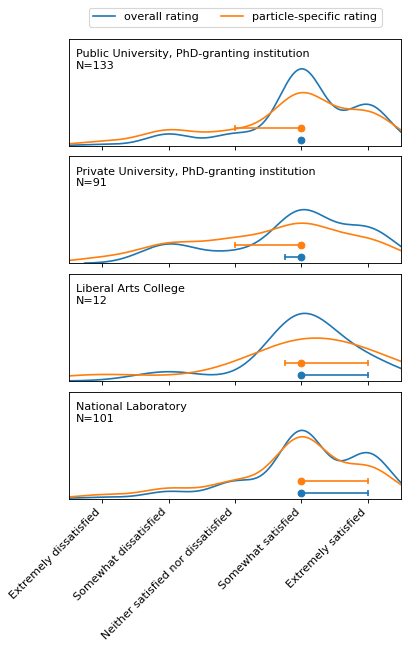}
    \caption{Satisfaction with graduate education as a function of current institutional affiliation. Overall satisfaction is shown in blue, particle-specific in orange. Quartiles are also presented. No statistically significant relationship was found.}
    \label{fig:GradSatisfactionByLocation}
\end{figure}

\begin{figure}[p!]
    \centering
    \includegraphics[width=\linewidth]{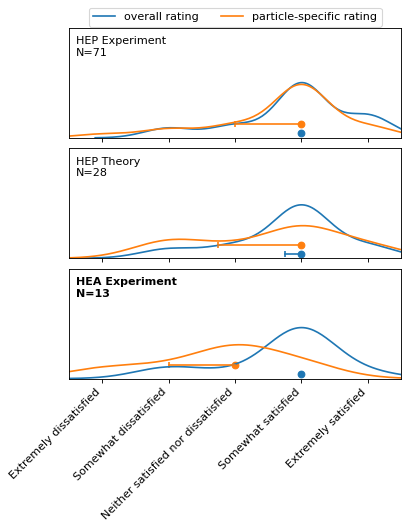}
    \caption{Satisfaction with graduate education as a function of current sub-field (graduate students and postdoctoral researchers only). Overall satisfaction is shown in blue, particle-specific in orange. Quartiles are also presented. The \textbf{bolded} categories are those for which there is a statistically significant relationship between categories and rating ($p < 0.05$ in a $\chi^2$ test of independence). HEA Theory not included due to small sample size. Further discussion of sub-field in Section~\ref{sec:subfieldprefs}.}
    \label{fig:GradSatisfactionBySubfield}.
\end{figure}

To understand perceptions of overall graduate education, all respondents were asked ``\textit{How satisfied do you feel with your graduate education in physics?}" and then ``\textit{How satisfied do you feel with your graduate education in \underline{particle} physics?}".

Ratings were generally positive across all current positions, with median ratings of `somewhat satisfied' across all respondents (Figure \ref{fig:GradSatisfactionOverall}). When broken down by position (Figure \ref{fig:GradSatisfactionByPosition}), tenured faculty were statistically more likely than other groups to rate their overall and particle-specific graduate education as `extremely satisfied' ($p=$ 0.01, $p<$ 0.01 respectively). Graduate students were statistically more likely than other groups to rate their overall graduate education as `somewhat satisfied' ($p<$ 0.01), but were also more likely than other groups to rate their particle-specific graduate education as `somewhat-dissatisfied' or `neither satisfied nor dissatisfied' ($p<$ 0.01).  

No relationship between current institution and satisfaction with overall or particle-specific graduate education was found (Figure \ref{fig:GradSatisfactionByLocation}).

When compared to current sub-field, the only significant relationship was that respondents who identified as working in 
`\textit{Gravity, astrophysics, and/or cosmology experiment}` were more likely than the other groups to rate their particle-specific education as less-than-satisfying ($p<$ 0.01), as seen in Figure \ref{fig:GradSatisfactionBySubfield}.

\subsection{Career Goals in Graduate School}
\label{sec:gradcareers}

\begin{figure}[p!]
    \centering
    \begin{subfigure}{.96\linewidth}
        \includegraphics[width=\linewidth]{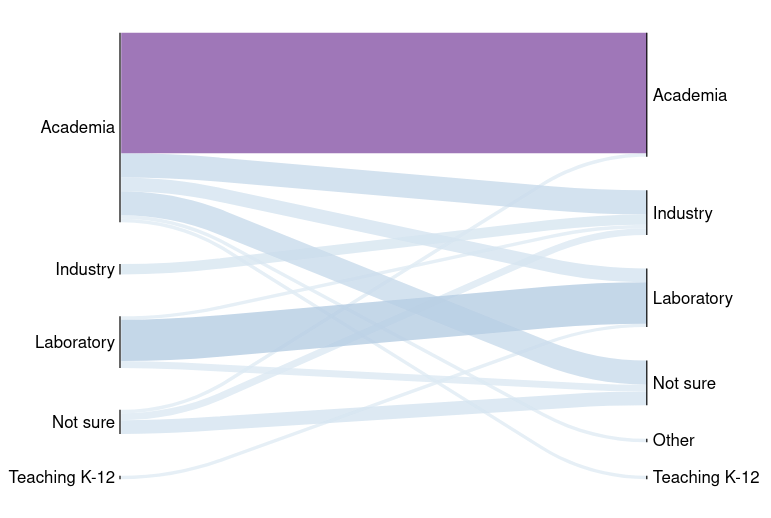}
        \caption{Graduate Students}
        \label{fig:CareerIntentGrad}
    \end{subfigure}
    \begin{subfigure}{.96\linewidth}
        \includegraphics[width=\linewidth]{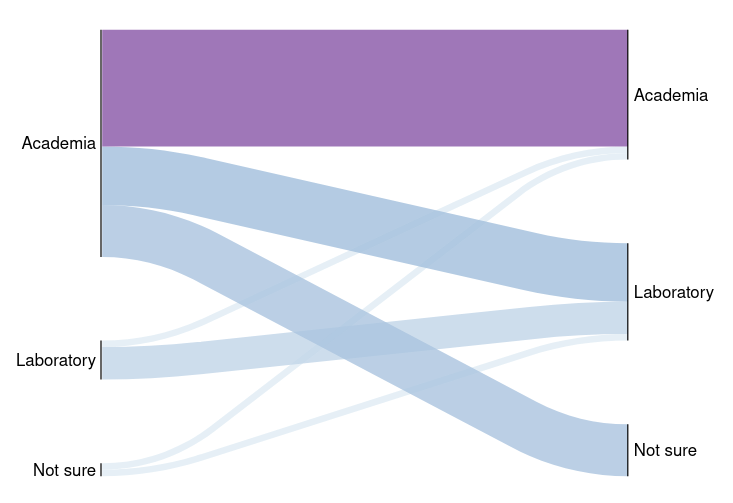}
        \caption{Postdoctoral Researchers}
        \label{fig:CareerIntentPostdoc}
    \end{subfigure}
    \begin{subfigure}{.96\linewidth}
        \includegraphics[width=\textwidth]{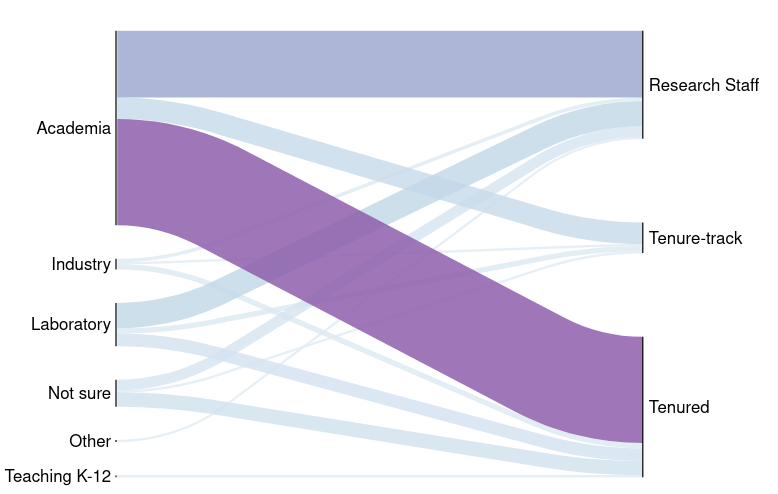}
        \caption{Tenure, Tenure-Track, and Research Staff}
        \label{fig:CareerIntentSenior}
    \end{subfigure}
    \caption{Flowcharts of ``\textit{career intended when starting graduate school}" (\textit{left}) vs ``\textit{current career intention}" (\textit{right}). Information is split between graduate student, postdoctoral researcher, and senior scholars (for the latter, their current position was considered their ``current career intention"). Note that the survey population suffers from survivorship bias, so we do not expect a significant number of ``industry" responses from postdoctoral researchers or senior scholars. }
    \label{fig:CareerIntentOverall}
\end{figure}

In an effort to identify the pipeline of respondents through their academic experience, all respondents were asked: ``\textit{What was your intended career goal when starting grad school?}", and junior respondents were asked ``\textit{What is your intended career goal now?}". Results were compiled into six categories: `Academia', `Industry', `Laboratory' (both national labs and private labs), `Teaching K-12', `Not Sure', and `Other' (with opportunity for text entry). For senior respondents, their career goal was assumed to be their current position --- see section \ref{sec:missingquestions} for comments on this choice. 

Flowcharts for junior and senior respondents are included as Figure \ref{fig:CareerIntentOverall}. The authors note that this survey suffers from survivorship bias: based on the style of distribution, both postdoctoral researchers and senior respondents are predominantly from academic or laboratory institutions, and therefore are less likely to respond that their career goal was or is industry.

\begin{figure*}[p!]
    \centering
    \begin{subfigure}{\linewidth}
        \begin{tabular}{p{0.5\linewidth} p{0.3\linewidth}}
        \hline
        \textbf{Survey Option} & \textbf{Results Label} \\
        \hline\hline
        High energy, nuclear, and/or particle theory & HEP Theory\\
        High energy, nuclear, and/or particle experiment & HEP Experiment\\
        Gravity, astrophysics, and/or cosmology theory & HEA Theory\\
        Gravity, astrophysics, and/or cosmology experiment & HEA Experiment\\
        Other (please describe) & Other \\
        \hline
        \end{tabular}
        \caption{Recategorization of Subfield Labels}
        \label{fig:SubfieldPrefLabels}
    \end{subfigure}
    \begin{subfigure}{.48\linewidth}
        \includegraphics[width=\linewidth]{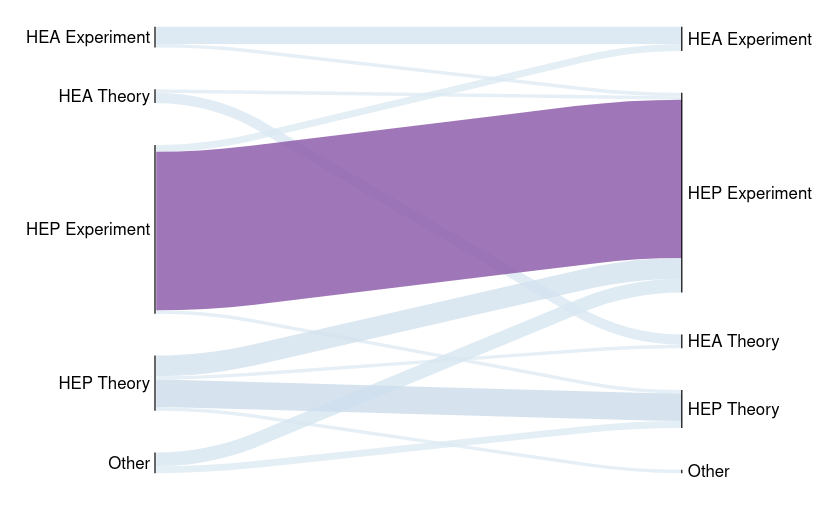}
        \caption{Graduate Students}
        \label{fig:SubfieldPrefGrad}
    \end{subfigure}
    \begin{subfigure}{.48\linewidth}
        \includegraphics[width=\linewidth]{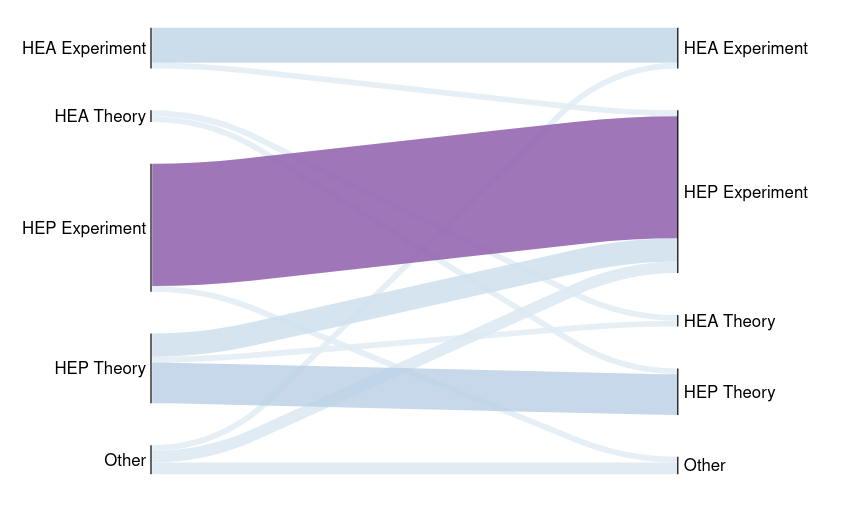}
        \caption{Postdoctoral Researchers}
        \label{fig:SubfieldPrefPostdoc}
    \end{subfigure}
    \caption{Flowcharts of ``\textit{sub-field intended when starting graduate school}" (\textit{left}) vs ``\textit{eventual sub-field}" (\textit{right}). Information is split between graduate student and postdoctoral researcher. Majority HEP Experiment is expected due to population of respondents, but we note the transition from theory to experiment that is not mirrored in the other direction.}
    \label{fig:SubfieldPrefOverall}
\end{figure*}

\clearpage 

Nevertheless, Figure \ref{fig:CareerIntentGrad} demonstrates that a not-insignificant portion of the graduate student respondents intend to pursue an industry career, which suggests that the skills that graduate students learn within the HEPA community are being exported to industry careers. Additionally, there is quite a bit of movement from `Academia' to both `Industry' and `Laboratory'. These findings are supported by statistics from the Snowmass 2021 Community Survey Report, which suggested that postdoctoral researchers were likely to apply for industry positions in STEM at least in part due to availability of industry jobs, work enviroment, and overall interest \cite{agarwalSnowmass2021Community2022}.

\subsection{Sub-field Preferences}
\label{sec:subfieldprefs}

Junior respondents (graduate students and postdocs) only were asked: ``\textit{When starting grad school, what was your intended area of research?}" and ``\textit{What is/was your eventual type of research in grad school?}". Available options are listed in Figure \ref{fig:SubfieldPrefLabels}. For those who responded `Other', a short textbox was provided. Additionally, participants were offered the opportunity to explain their sub-field switch, if they desired to. 
129 participants responded to these questions. Flowcharts derived from these results are presented in Figure \ref{fig:SubfieldPrefOverall}.

About one quarter of the respondents reported that they changed their sub-field during graduate school. Of the respondents who offered a reason for their field switch, two thirds discussed the lack of positions and/or lack of funding in their desired sub-field, with specific discussion of the lack of funding for HEPA Theory. A further third mentioned their advisor was the reason for the switch: either moving away from a poor advisor, or towards a better one. We also note that switches from theory to experiment were much more common than from experiment to theory (both within HEA and HEP, and between the two).

The respondents who responded `Other' for preliminary subfield responded that they switched to HEPA from the following physics subfields: condensed matter (54\%), biophysics (18\%), and undeclared (18\%).

\subsection{Career Preparedness}
\label{sec:careerprep}

\begin{figure}[h!]
    \centering
    \includegraphics[width=\linewidth]{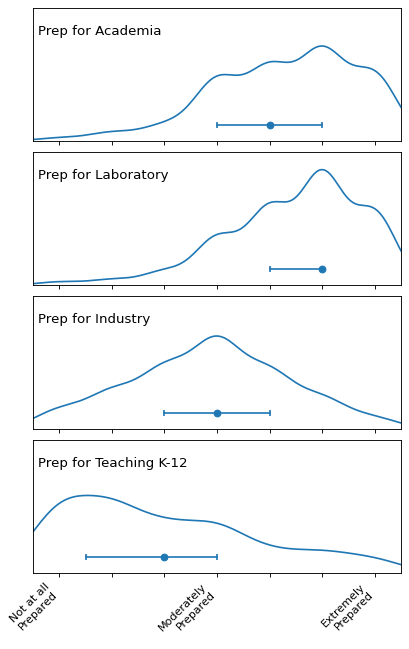}
    \caption{Preparation for given career across all respondents. When compared to preparation for academia or laboratory careers, respondents rated their preparation for industry and teaching K-12 significantly lower ($p<0.01$). Quartiles are also presented. Further discussion of preparedness in section \ref{sec:careerprep}.}
    \label{fig:CareerPrep_Overall}.
\end{figure}

\begin{figure}[p!]
    \centering
    \includegraphics[width=\linewidth]{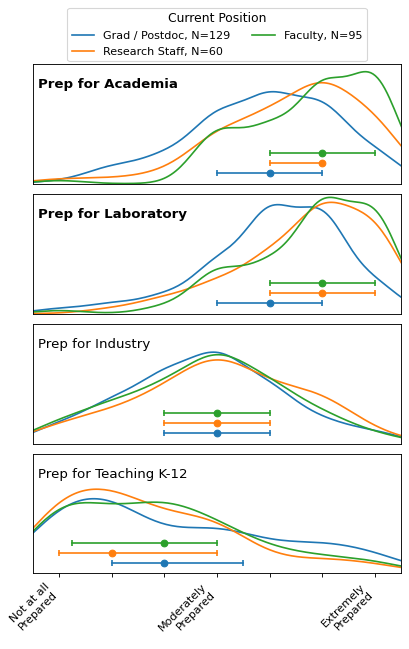}
    \caption{Preparation for given career as a function of current affiliation. Quartiles are also presented. The \textbf{bolded} categories are those for which there is a statistically significant relationship between categories and rating ($p < 0.05$ in a $\chi^2$ test of independence).  Further discussion of preparedness in section \ref{sec:careerprep}.}
    \label{fig:CareerPrep_Position}.
\end{figure}

\begin{figure}[p!]
    \centering
    \includegraphics[width=\linewidth]{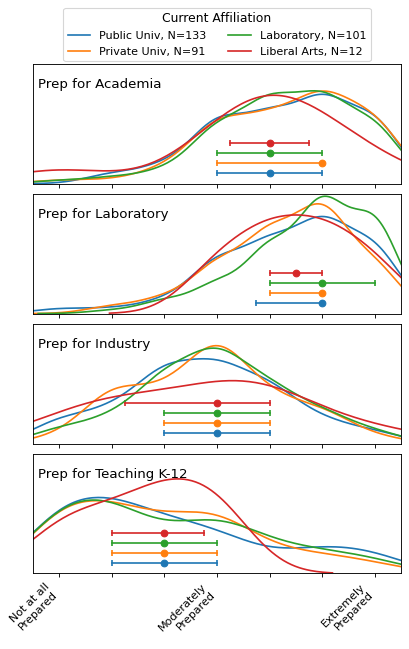}
    \caption{Preparation for given career as a function of current affiliation. Quartiles are also presented. No statistically significant relationships were found. Further discussion of preparedness in section \ref{sec:careerprep}.}
    \label{fig:CareerPrep_Location}.
\end{figure}

Respondents were asked ``\textit{On a scale of 1 to 7, how well do you feel graduate school prepared you for the following career paths: College or University Position, National/Private Laboratory Scientist, Industry Job, Teaching K-12?}". Data for all respondents can be found in Figure \ref{fig:CareerPrep_Overall}. Respondents percieved themselves as moderately prepared for careers in academia, laboratories, and industry, but felt unprepared for teaching K-12.

\begin{figure}[h!]
    \centering
    \includegraphics[width=\linewidth]{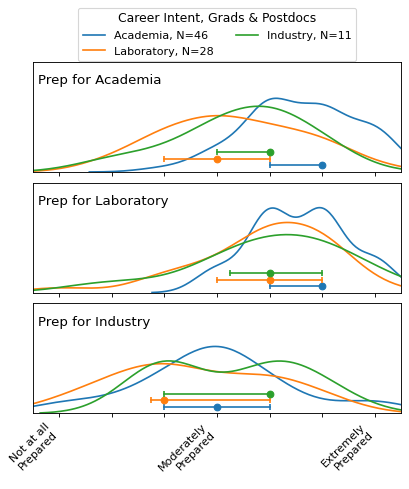}
    \caption{Preparation for given career as a function of intended career path (graduate students and postdoctoral researchers only). Quartiles are also presented. No statistically significant relationships could be reported (and data for those intending to pursue K-12 teaching are not reported) due to small sample size. Further discussion of preparedness in section \ref{sec:careerprep}.}
    \label{fig:CareerPrep_CareerGoal}.
\end{figure}

Responses were separated by current position, current affiliation, and intended career path (the latter only consisting of graduate students and postdocs). These data can be found in Figures \ref{fig:CareerPrep_Position}, \ref{fig:CareerPrep_Location}, and \ref{fig:CareerPrep_CareerGoal}, respectively. When compared to preparation for academia or laboratory careers, respondents rated their preparation for industry and teaching K-12 significantly lower ($p<0.01$). No statistically significant relationship was found between preparation and current affiliation. No statistically significant relationship between career goal and preparation for that career could be reported due to small sample size.

\clearpage 

When analyzed as a function of position, preparation for academia was rated significantly higher for faculty (tenure-track and tenured) ($p<0.01$) and significantly lower for graduate students and postdocs ($p<0.01$) when compared with other groups. Preparation for laboratory positions had the same relationships, but only the lower rating from graduate students was statistically significant ($p<0.01$). Interestingly, Research Staff did not feel significantly more prepared for their current position compared to other groups ($p>0.9$). All groups felt equally ``moderately prepared" for careers in industry.

Although we could not report statistically significant relationships for preparation as a function of career intent (due to small sample size), there are some interesting trends. Graduate students and postdocs who intended to pursue a career in academia reported that they felt almost equally prepared for academic and laboratory jobs. Those who wanted to pursue a career at a laboratory or in industry reported that they were prepared for laboratory jobs, but not for industry. Finally, junior respondents felt only ``moderately prepared" for a career in industry, even if it was their chosen career path.

\subsection{Professional, Experimental, and Theoretical Skills}
\label{sec:gradskills}

Survey respondents were shown three sets of skills that one might expect to use in a particle physics career: nine professional skills, six technical skills most likely to be used by experimentalists, and five mathematical skills most likely to be used by theorists. The skills are listed in the Appendix and most plots in this section. For each skill, respondents were asked: \textit{``For each of these topics, please state if you received any experience in it during your own undergraduate + graduate education.''} This question was mandatory.

If the respondent answered ``Yes'', they were asked two further questions. First, \textit{``For each of these topics that you did gain experience with during undergraduate + graduate school, what was the primary mode of training you received? (Select the last option `self-taught' if you received no training.)''} The possible responses were:

\begin{itemize}\setlength\itemsep{0em}
\item In a formal course at your university
\item Online, through a structured program (such as edX or Coursera)
\item In a course organized outside of the classroom (e.g. summer school, CERN courses, professional workshop, formal training in your research group, etc.)
\item Mentoring or peer learning
\item As needed; in a decentralized manner, self-taught
\end{itemize}

\noindent Second, \textit{``Please rate the method of training you selected above.''} The possible responses were:
\begin{itemize}\setlength\itemsep{0em}
\item Extremely dissatisfied
\item Somewhat dissatisfied
\item Neither satisfied nor dissatisfied
\item Somewhat satisfied
\item Extremely satisfied
\end{itemize}

If the respondent answered ``No'' to receiving experience in their schooling, they were asked: \textit{``For each of these topics that you did not experience during your undergraduate + graduate education, how important has this topic been in your career so far?''} The possible responses were:
\begin{itemize}\setlength\itemsep{0em}
\item Not at all important
\item Slightly important
\item Moderately important
\item Very important
\item Extremely important
\end{itemize}

\subsubsection{Professional Skills}
\label{sec:profskills}

\begin{figure}
{\includegraphics[width=3.4in]{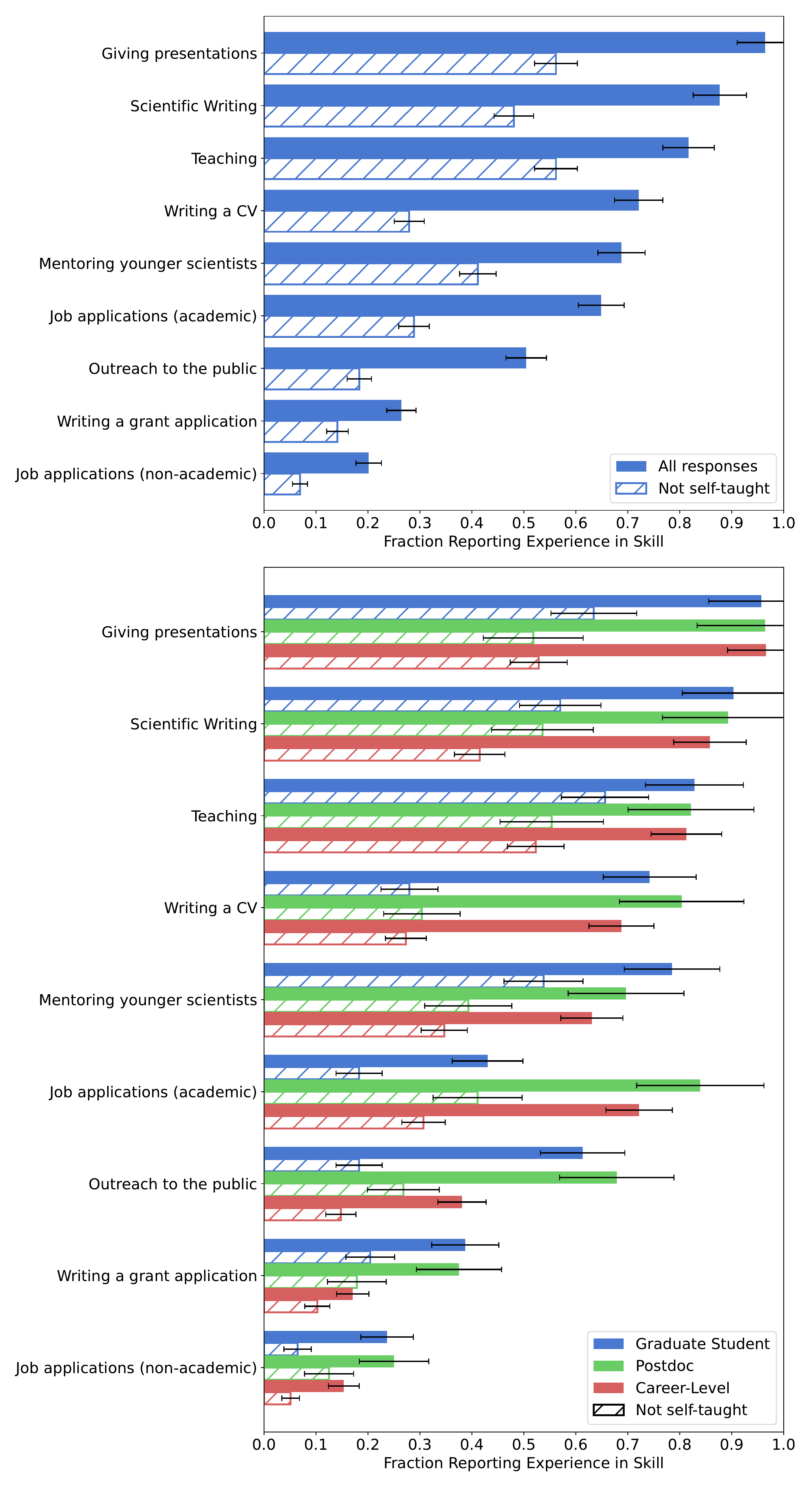}}
\caption{Responses to the question: \textit{``For each of these topics, please state if you received any experience in it during your own undergraduate + graduate education.''} \textbf{(Top)} Solid-shaded bars show the fraction of all respondents reporting that they received each professional skill during their own education. Hatch-shaded bars show this quantity with an extra restriction that respondents' selected mode of training was not ``As needed; in a decentralized manner, self-taught.'' Statistical Poisson errors are shown, calculated ($\sqrt{\text{N~``Yes''~responses}}$~/~N~responses). \textbf{(Bottom)} The fraction of respondents in each job category reporting the same. ``Career-Level'' includes research staff, nontenured, and tenured faculty. The bars for each skill are stacked in the same vertical order as in the legend.}
\label{fig:Professional_Experience}
\end{figure}

\begin{figure}
{\includegraphics[width=3.4in]{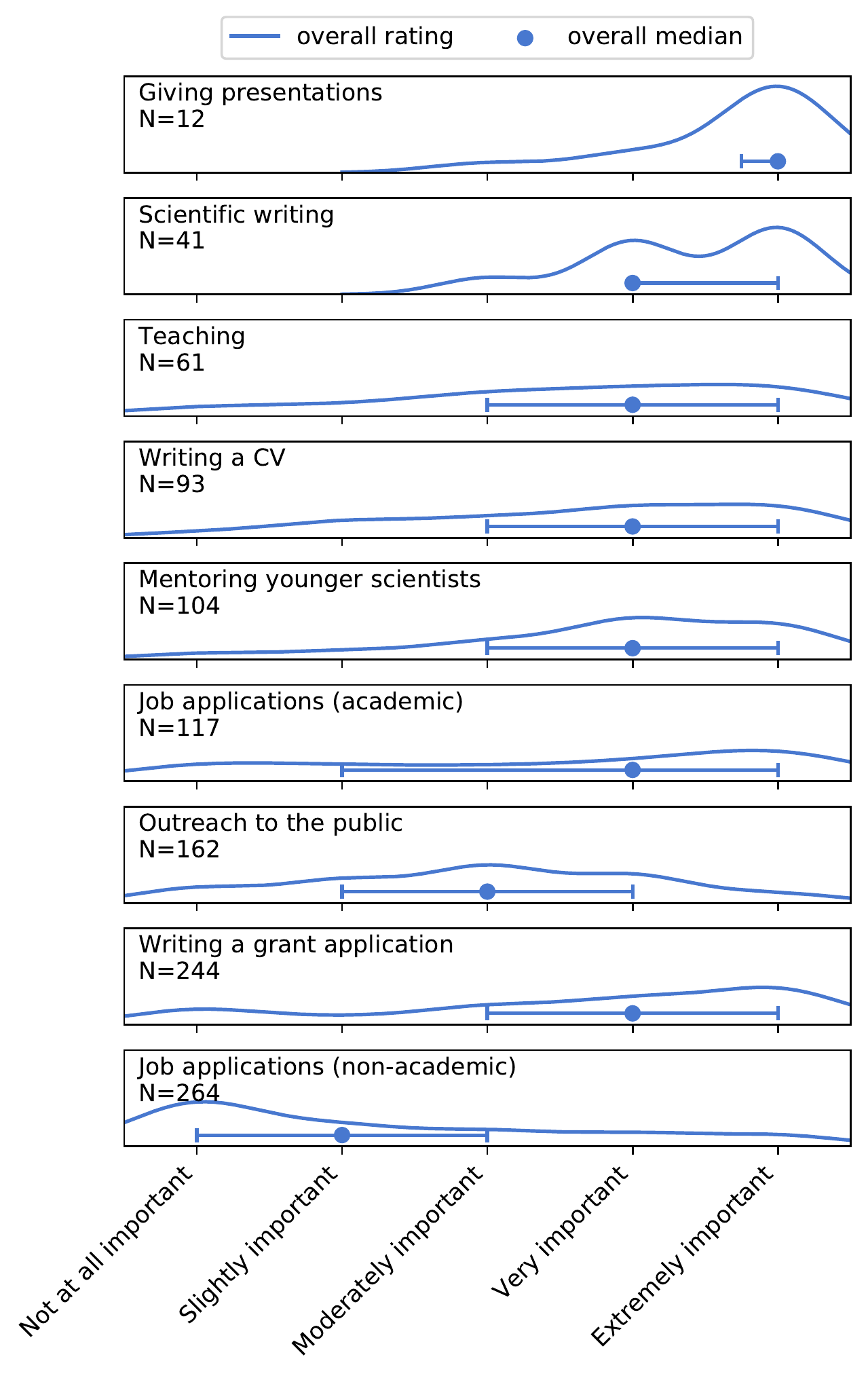}}
\caption{The distribution of respondents selecting each rating of importance to their career so far for each professional skill. Height is proportional to the number of responses with a given rating. }
\label{fig:Violin_Importance_Prof}
\end{figure}

\begin{figure}
{\includegraphics[width=3.4in]{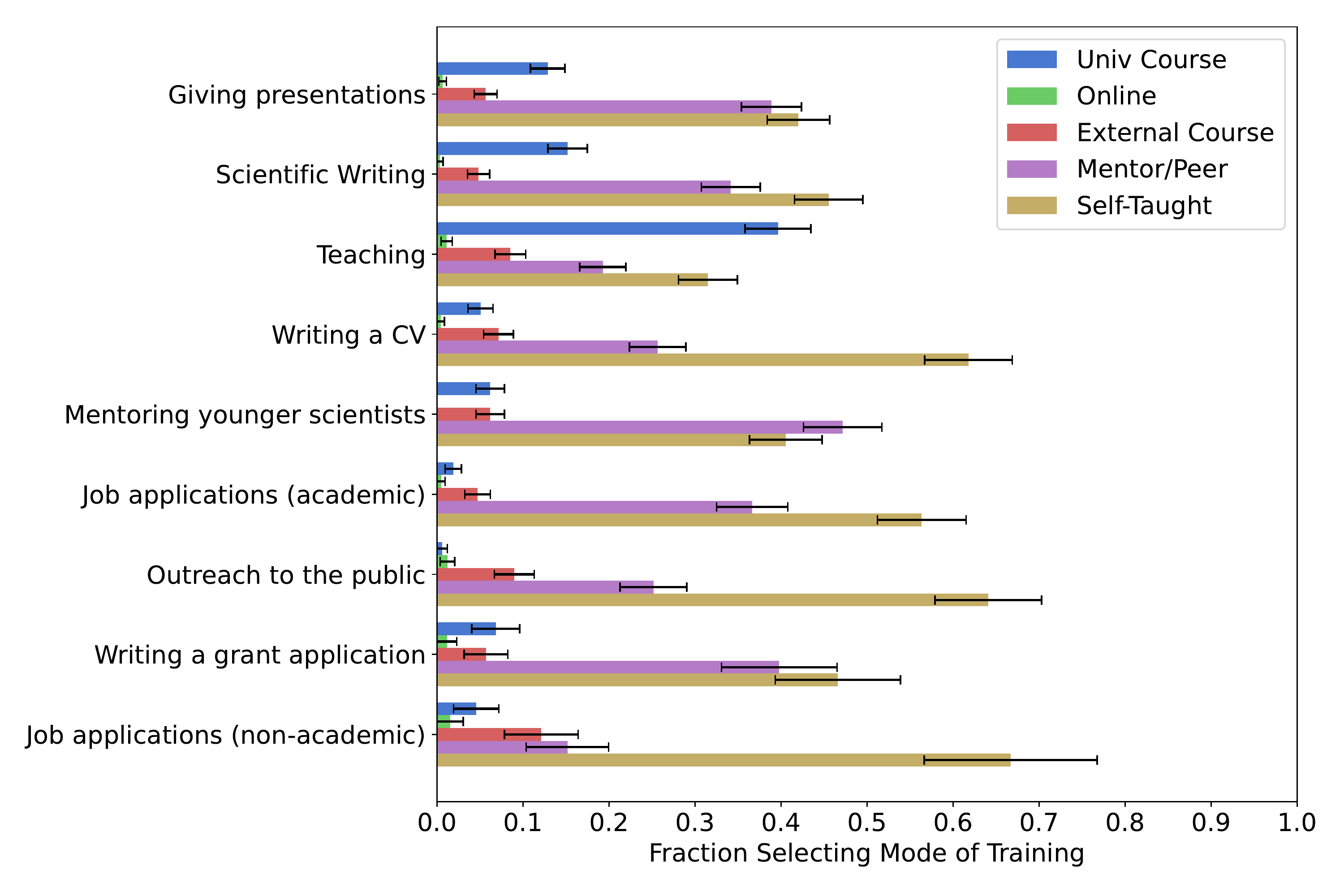}}
\caption{Responses to the question: \textit{``For each of these topics that you did gain experience with during undergraduate + graduate school, what was the primary mode of training you received? (Select the last option `self-taught' if you received no training.)''} For each skill, we display the fraction of respondents selecting each mode out of all respondents who chose a mode. This question was not mandatory, so a few respondents said they gained experience in a skill, but did not select a modality; these responses are not included in the denominator. The five results for each skill add up to 1. Statistical Poisson errors are shown, calculated ($\sqrt{\text{N~``Mode''~responses}}$~/~N~``Any~Mode''~responses). ``Career-Level'' includes research staff, nontenured, and tenured faculty. The bars for each skill are stacked in the same vertical order as in the legend. The legend is abbreviated; see Section~\ref{sec:gradskills} for the exact wording of the modes.}
\label{fig:Professional_Mode}
\end{figure}

\begin{figure}
{\includegraphics[width=3.4in]{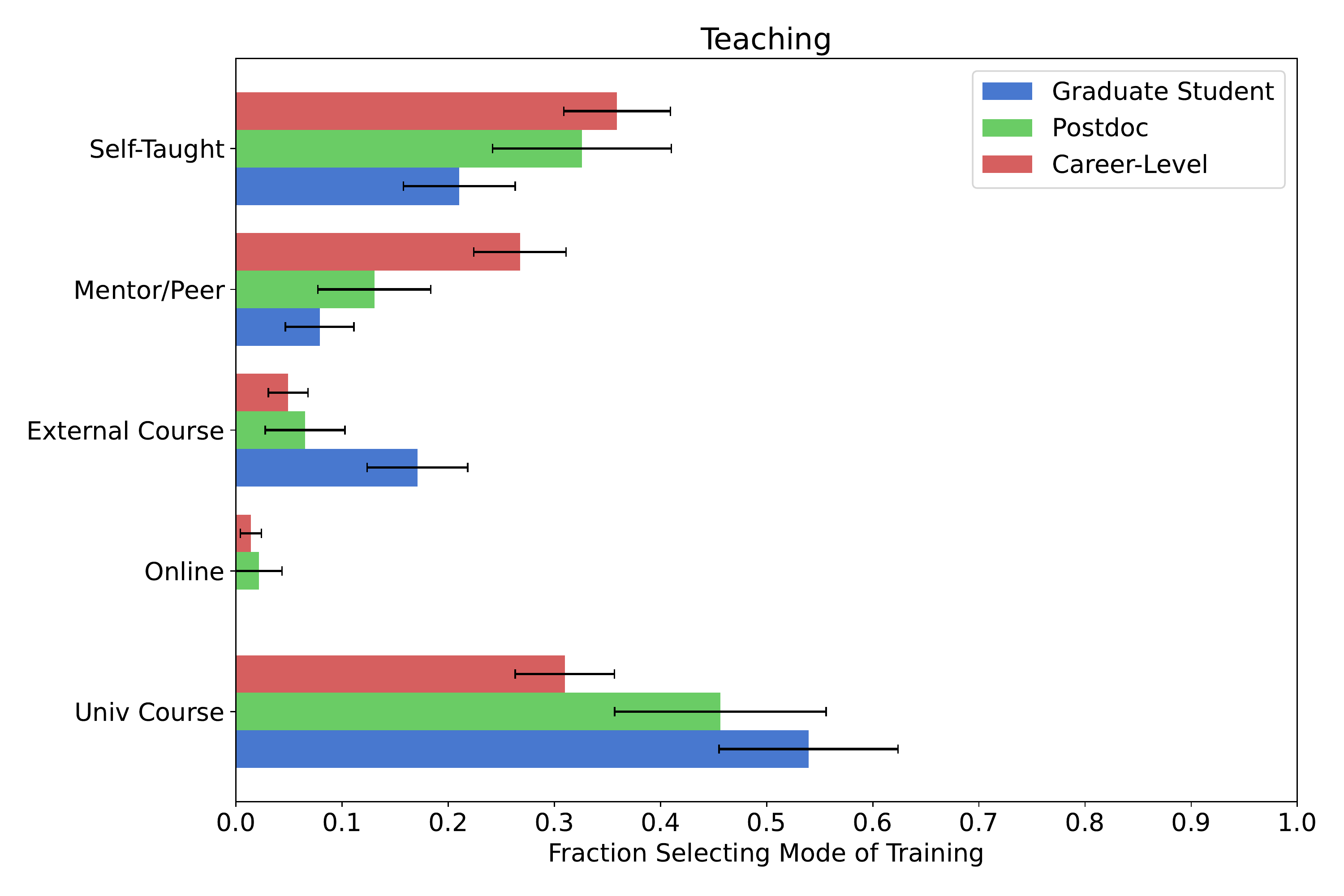}}
\caption{The fraction of respondents in each job category selecting each mode of training for Teaching, out of all respondents in that job category who chose a mode of training for their Teaching experience. Respondents who said they gained experience in Teaching but did not select a mode are removed. Thus, the five bars for each job category add up to 1. Statistical Poisson errors are shown, calculated ($\sqrt{\text{N~MODE~responses}}$~/~N~ANY~MODE~responses). The bars for each mode are stacked in the same vertical order as in the legend. For this skill, we see a moderate correlation between career position and formality of training; younger scientists are more likely to be trained in some type of course. See Table~\ref{tab:age_dependent_skills} for a list of all skills with a correlation between age and mode of training.}
\label{fig:Professional_Mode_By_Career_2}
\end{figure}

In the professional skills section, we asked about a variety of professional skills that are critical to a particle physics career, as well as to other careers. These are listed in the top panel of Figure~\ref{fig:Professional_Experience}, along with the percentage of respondents who received any experience in each skill during their undergraduate + graduate education. Most respondents gained experience with most skills. One optimistic takeaway is that the most-experienced skills align with the skills that respondents found important, illustrated in Figure~\ref{fig:Violin_Importance_Prof}.

\begin{table*}[t!]
    \begin{tabular}{p{2.5in} p{1.36in} p{1.36in} p{1.36in}}
        \begin{tabular}[c]{@{}l@{}}Skill\end{tabular} & 
        \begin{tabular}[c]{@{}l@{}}Number of \\responses\end{tabular} & 
        \begin{tabular}[c]{@{}l@{}}p-value\end{tabular} & 
        \begin{tabular}[c]{@{}l@{}}Correlation with \\formal training ($\tau$) \end{tabular} \\ \hline\hline\\
        Mentoring young scientists & 227 & 0.032 &-0.15\\
        \textbf{Teaching} & 270 & $2.0\times10^{-5}$ & -0.23\\
        Scientific writing & 290 & 0.0025 & -0.17\\
        Giving presentations & 319 & 0.0018 & -0.13\\
        \textbf{Statistics} & 256 & 0.0029 & -0.20\\
        \textbf{Electrical engineering} & 126 & 0.025 & -0.26\\ 
        \textbf{Mechanical engineering} & 72 & 0.041 & -0.23 \\ 
        \textbf{Formal math} & 138 & 0.022 & +0.20 \\ \\ \hline
    \end{tabular}
\caption{A list of skills in which the received mode of training was significantly ($p<0.05$) correlated with with career position, according to a $\chi^2$ test. We consider self-taught to be the least formal method, mentoring or peer learning to be more formal, and all other modes such as a university course to be most formal for the purposes of ranking the categories. Negative values of $\tau$ correspond to skills that were more likely to be taught in a formal manner (as opposed to self-taught or through a peer) for respondents currently in earlier career stages. This suggests a transition over time to more formal training for that skill. We consider correlations with $0.1<|\tau|\le0.2$ to be weak, and those with $0.2<|\tau|\le0.3$ to be moderate, highlighted in \textbf{bold}. An example skill with a moderately strong correlation was found is given in Figure~\ref{fig:Professional_Mode_By_Career_2}.}
\label{tab:age_dependent_skills}
\end{table*}

\begin{table*}
    \begin{tabular}{p{2.04in} p{2.04in} p{0.90in} p{0.90in} p{0.90in}}
        \begin{tabular}[c]{@{}l@{}}Skill\end{tabular} &   \begin{tabular}[c]{@{}l@{}}Mode\end{tabular} &                          \begin{tabular}[c]{@{}l@{}}High or low\\rating?\end{tabular} &  
        \begin{tabular}[c]{@{}l@{}}Number of\\responses\end{tabular} &  \begin{tabular}[c]{@{}l@{}}p-value\end{tabular} \\ \hline\hline\\
        \textit{Mentoring young scientists} & As needed; self-taught & Low & 90 & $9.5\times10^{-6}$\\
        \textit{Teaching} & As needed; self-taught & Low & 84 & $7.1\times10^{-5}$\\
        \textit{Scientific writing} & As needed; self-taught & Low & 131 & $3.8\times10^{-4}$\\
        \textit{Giving presentations} & As needed; self-taught & Low & 133 & $6.3\times10^{-6}$\\
        \textit{Outreach to the public} & As needed; self-taught & Low & 106 & 0.031\\
        \textit{Writing a CV} & As needed; self-taught & Low & 145 & $2.0\times10^{-4}$\\
        \textit{Job applications (academic)} & As needed; self-taught & Low & 119 & $5.8\times10^{-4}$\\
        \textbf{Statistics} & As needed; self-taught & Low & 63 & $9.1\times10^{-6}$\\
        Theory-oriented particle physics & As needed; self-taught & Low & 15 & 0.019\\
        Mathematics for formal theories & As needed; self-taught & Low & 11 & 0.0041\\
        \\\hline\\
        \textit{Mentoring young scientists} & Mentoring or peer learning & High & 105 & 0.011\\
        \textit{Giving presentations} & Mentoring or peer learning & High & 124 & 0.015\\
        \textit{Job applications (academic)} & Mentoring or peer learning & Low & 78 & 0.013\\
        \textbf{Statistics} & Mentoring or peer learning & Low & 31 & $2.7\times10^{-4}$\\
        Mathematics for formal theories & Mentoring or peer learning & Low & 6 & 0.014\\
        Gravity, GR, and cosmology & Mentoring or peer learning & Low & 7 & 0.0028\\
        \\ \hline \\
        \textit{Scientific writing} & External course & High & 14 & 0.038 \\
        \textit{Giving presentations} & External course & High & 18 & 0.011\\
        \\ \hline \\
        \textit{Teaching} & University course & High & 105 & $1.6\times10^{-4}$\\
        \textit{Scientific writing} & University course & High & 43 & 0.0077 \\
        \textit{Writing a CV} & University course & High & 12 & 0.04\\
        \textbf{Statistics} & University course & High & 139 & 0.030 \\
        Theory-oriented particle physics & University course & High & 166 & 0.011\\
        Mathematics for formal theories & University course & High & 116 & 0.0089\\ \\ \hline
    \end{tabular}
\caption{A list of skills for which the mode of training is significantly ($p<0.05$) correlated with a respondent's rating of that mode, according to a $\chi^2$ test of independence. For each skill, the ratings were checked by eye to determine whether the rating was high or low, relative to the null hypothesis of mode-rating independence---for example, see Figure~\ref{fig:Professional_Mode_By_Career_2}. We also display the number of responses in each skill-mode pairing and the p-value obtained from the $\chi^2$ test. In the leftmost column, professional skills are \textit{italicized}, experimental skills are \textbf{bolded}, and theoretical skills are in plain text.}
\label{tab:pvalues}
\end{table*}

However, when examining the mode of training, the results are more humbling. We show all possible responses and their corresponding rates in Figure~\ref{fig:Professional_Mode}; in the hatched bars of Figure~\ref{fig:Professional_Experience}, we show the percentage of respondents who gained experience and selected a mode of training other than ``Self-Taught.'' Strikingly, these skills are primarily being conveyed through ad-hoc self-training or peer mentorship, rather than through formal avenues. Apart from Teaching (which was at a still-low 40\%), less than 20\% of respondents who gained experience in a skill reported that they were trained through a university course. Barely any professional skills we tested had a majority of respondents stating that they received actual training in the subject. Even for critical academic skills like giving presentations and teaching, the fraction was barely above 50\%. In other words---respondents are gaining the skills they find important, but they are forced to figure it out themselves or rely on their peers and mentors for informal transfers of knowledge.

There is evidence that, for some skills, physicists are being trained through more formal mechanisms over time. See the bottom panel of Figure~\ref{fig:Professional_Experience}, similar to the top panel but broken out by career position. For some skills, there is a trend that scientists in later career stages were more likely to gain training through self-teaching or peer mentorship, and less likely through coursework. Teaching is a good example, exhibited in Figure~\ref{fig:Professional_Mode_By_Career_2}. For the ``Self-Taught'', ``Mentor/Peer'', and ``Univ Course'' categories,  there is a clear career-stage-based trend. To confirm this trend, in Table~\ref{tab:age_dependent_skills}, we show a list of skills for which a $\chi^2$ test of independence confirms this relationship at a significance of $p<0.05$. Half of the professional skills show this trend, although the correlations are weak to moderate.

\begin{figure}
{\includegraphics[width=3.4in]{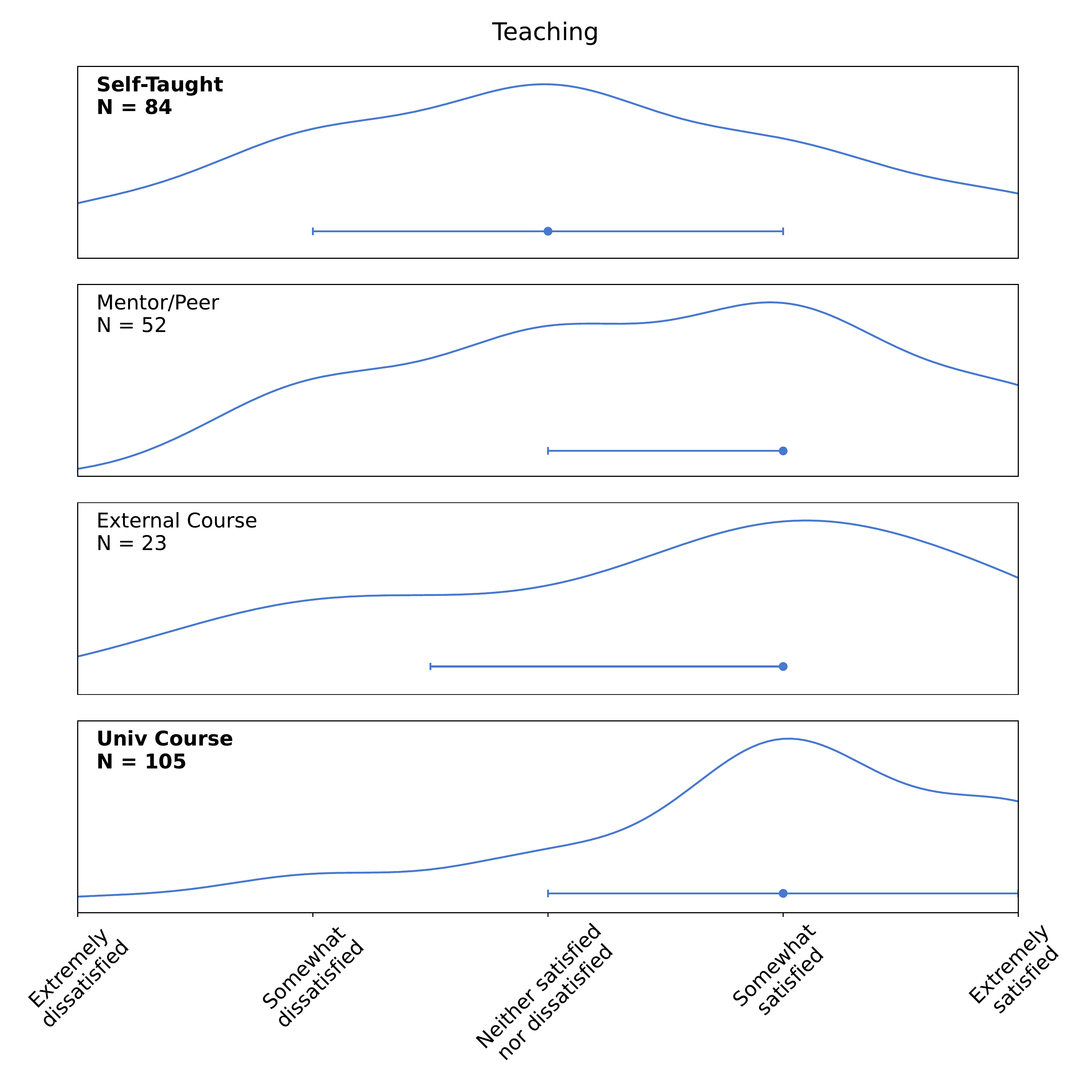}}
\caption{Distribution of responses to the question \textit{``Please rate the method of training you selected above''} for Teaching. Height is proportional to the fraction of responses with a given rating, for each mode. The horizontal line shows the median (dot) and 25\% and 75\% quantiles (vertical endcap). The \textbf{bolded modes} are those for which there is a statistically significant relationship between mode and rating ($p<0.05$ in a $\chi^2$ test). In this example, ``Self-Taught'' respondents had significantly lower ratings, and ``University Course'' respondents had significantly higher ratings. Only three respondents were trained in an online course, so these are omitted. See Table~\ref{tab:pvalues} for a full list of skills with a significant mode-rating relationship.}
\label{fig:HalfViolin_Rating_Prof_2}
\end{figure}

The fact that professional skills are generally passed through informal methods is not, by itself, concerning. However, when paired with self-reported satisfaction of different training modes, a clear problem arises. We performed a $\chi^2$ test of independence between training mode and training satisfaction for all the tested skills. The results are shown in Table~\ref{tab:pvalues}. If the test returned an unusually high $\chi^2$ (equivalent to an unusually low p-value), this indicated that a significantly high fraction of respondents were satisfied or dissatisfied with that mode of training. We checked the distributions manually to determine which (see Figure~\ref{fig:HalfViolin_Rating_Prof_2} for an example).

For professional skills---and indeed for all skills---the ``Self-Taught'' mode was broadly disfavored. It was given a significant low rating compared to the other modes, for every skill that returned $p<0.05$ on the $\chi^2$ test. This included seven of the nine professional skills. Meanwhile, when respondents reported an ``External course'' or ``University course'' mode, they were far more likely to rate this mode highly. Many professional skills did not exhibit a statistically significant result, mainly due to lack of statistics---remember that few respondents gained these skills through a formal method. However, when these modes were significant, they were universally given high ratings. The ``Mentoring and peer learning'' mode was more mixed. Two professional skills, \textit{Mentoring Young Scientists} and \textit{Giving Presentations}, were given high ratings through this mode, while \textit{Academic Job Applications} was given a low rating.

\begin{figure}
{\includegraphics[width=3.4in]{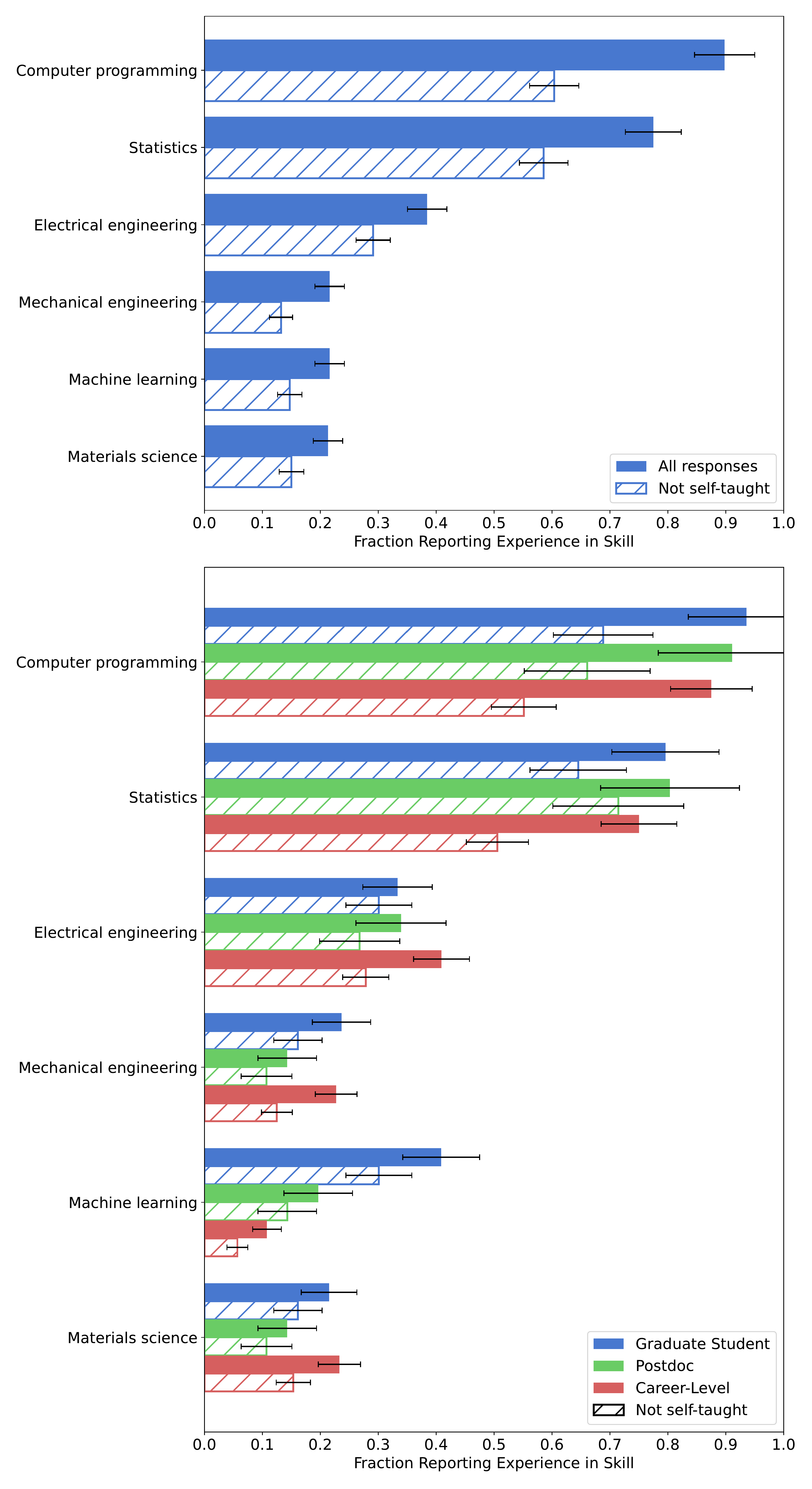}}
\caption{Responses to the question: \textit{``For each of these topics, please state if you received any experience in it during your own undergraduate + graduate education.''} \textbf{(Top)} Solid-shaded bars show the fraction of all respondents reporting that they received each technical skill during their own education. Hatch-shaded bars show this quantity with an extra restriction that respondents' selected mode of training was not ``As needed; in a decentralized manner, self-taught.'' Statistical Poisson errors are shown, calculated ($\sqrt{\text{N~``Yes''~responses}}$~/~N~responses). \textbf{(Bottom)} The fraction of respondents in each job category reporting the same. ``Career-Level'' includes research staff, nontenured, and tenured faculty. The bars for each skill are stacked in the same vertical order as in the legend.}
\label{fig:Technical_Experience}
\end{figure}

The story for professional skills is thus: people are getting some experience in the skills they need, but the mode of gaining these skills is not ideal. Respondents want formal training through some type of coursework, but they are usually unable to receive it. There is some evidence that this situation is improving over time, but only moderately.

\subsubsection{Technical Skills}
\label{sec:techskills}

\begin{figure}
{\includegraphics[width=3.4in]{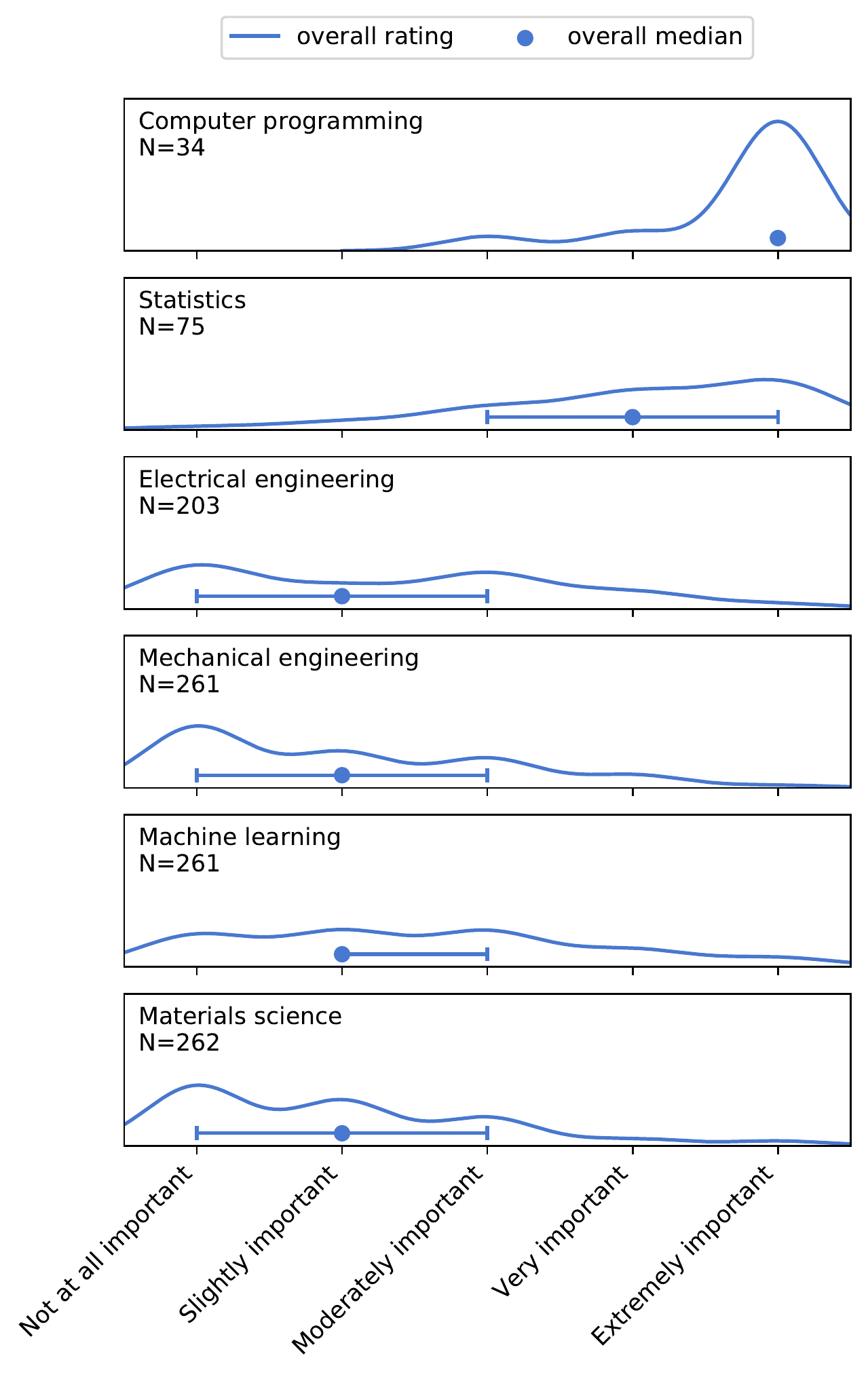}}
\caption{The distribution of respondents selecting each rating of importance to their career so far for each experimental technical skill, from among those who did not receive the skill. Height is proportional to the number of responses with a given rating. }
\label{fig:Violin_Importance_Tech}
\end{figure}

\begin{figure}
{\includegraphics[width=3.4in]{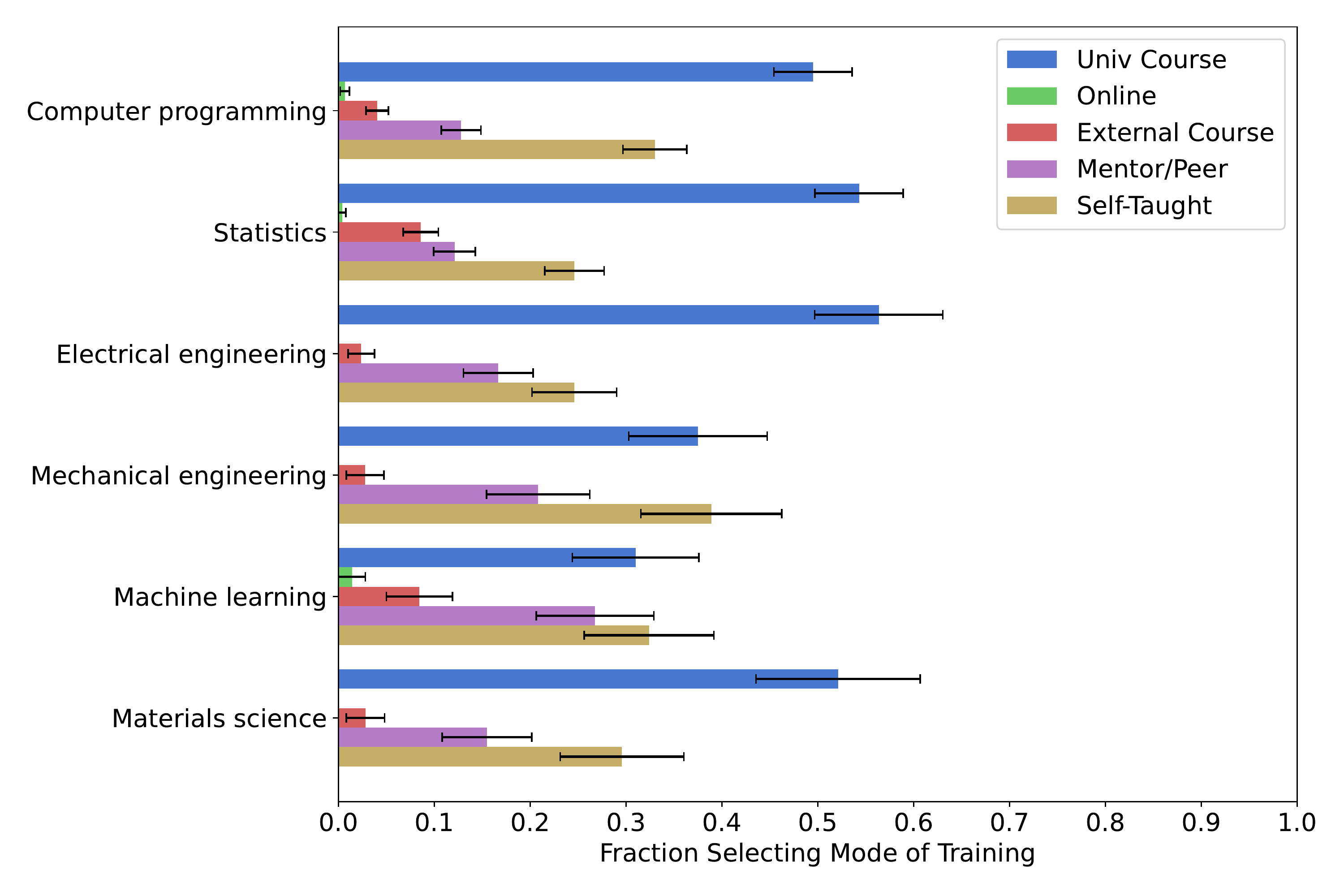}}
\caption{For each skill, we display the fraction of respondents selecting each mode out of all respondents who chose a mode. This question was not mandatory, so a few respondents said they gained experience in a skill, but did not select a modality; these responses are not included in the denominator. The five results for each skill add up to 1. Statistical Poisson errors are shown, calculated ($\sqrt{\text{N~``Mode''~responses}}$~/~N~``Any~Mode''~responses). ``Career-Level'' includes research staff, nontenured, and tenured faculty. The bars for each skill are stacked in the same vertical order as in the legend. The legend is abbreviated; see Section~\ref{sec:gradskills} for the exact wording of the modes.}
\label{fig:Technical_Mode}
\end{figure}

In the technical skills section, we asked about a variety of skills that are critical to a particle physics career, primarily (but not exclusively) to experimentalists. These are listed in the top panel of Figure~\ref{fig:Technical_Experience}, along with the percentage of respondents who received any experience in each skill. In general, the skills obtained tend to align with how important they were rated Figure~\ref{fig:Violin_Importance_Tech}. Only two skills---computer programming and statistics---displayed a majority of respondents receiving experience in those skills. These were the same two skills rated as most important in a particle physics career by the respondents who had not received these skills.

When we examine Figure~\ref{fig:Technical_Mode}, we can see that the mode of training is more mixed than for professional skills. For most technical skills, the plurality training mode is a university course. Nevertheless, a significant fraction of respondents gained these skills through self-teaching or peer mentorship---usually a majority of respondents, with those two modes combined. Similarly to the professional skills, Table~\ref{tab:age_dependent_skills} shows that for some technical skills, this pattern has changed over time. Specifically, for statistics, electrical engineering, and mechanical engineering, there is a moderate negative correlation between career stage and formality of training. Scientists currently in earlier career stages are more likely to have gained experience in these skills through some type of coursework as a graduate student.

\begin{figure}
{\includegraphics[width=3.4in]{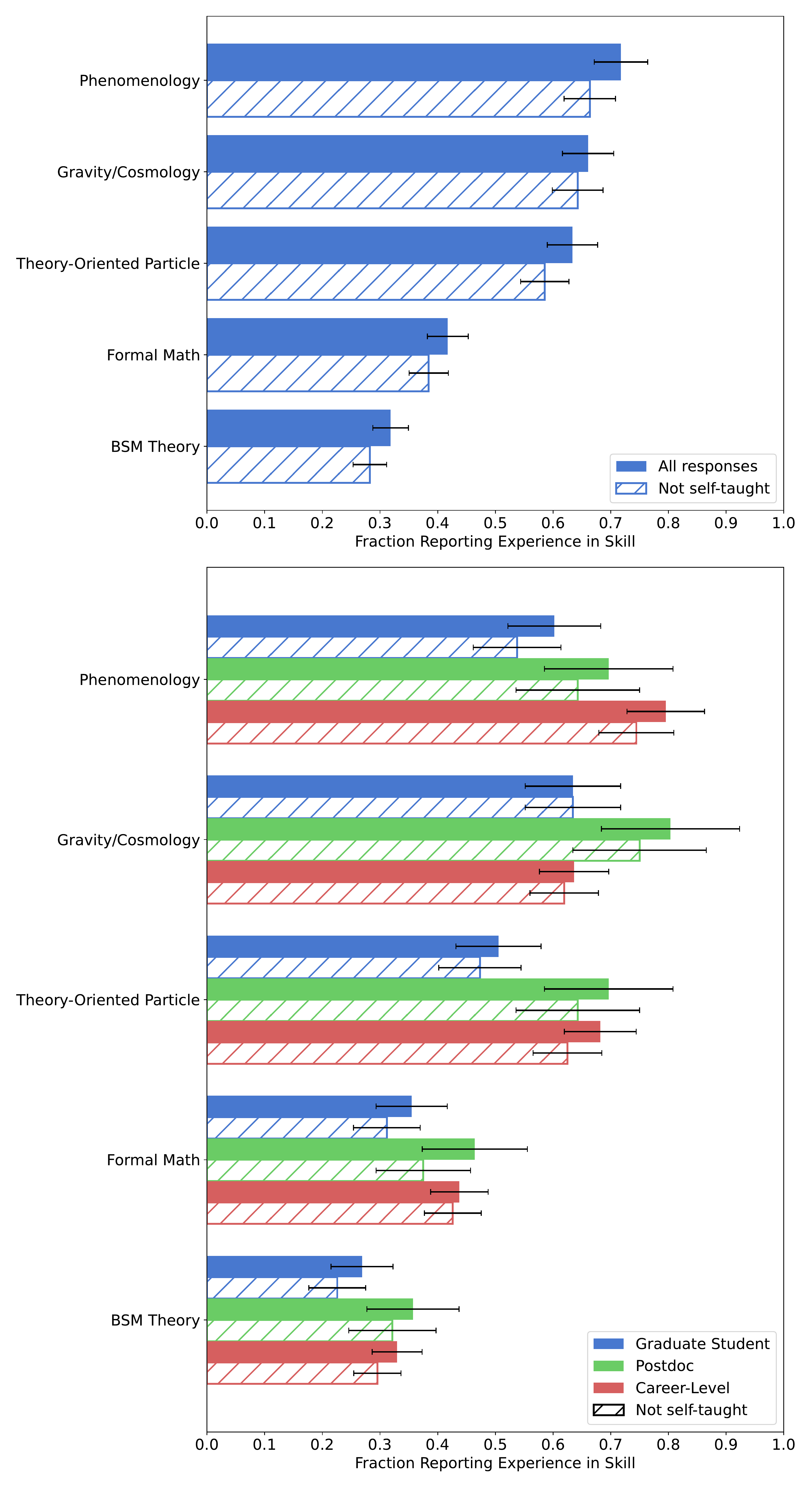}}
\caption{Responses to the question: \textit{``For each of these topics, please state if you received any experience in it during your own undergraduate + graduate education.''} \textbf{(Top)} Solid-shaded bars show the fraction of all respondents reporting that they received each theoretical mathematical skill during their own education. Hatch-shaded bars show this quantity with an extra restriction that respondents' selected mode of training was not ``As needed; in a decentralized manner, self-taught.'' Statistical Poisson errors are shown, calculated ($\sqrt{\text{N~``Yes''~responses}}$~/~N~responses). \textbf{(Bottom)} The fraction of respondents in each job category reporting the same. ``Career-Level'' includes research staff, nontenured, and tenured faculty. The bars for each skill are stacked in the same vertical order as in the legend.}
\label{fig:Mathematical_Experience}
\end{figure}

Respondents' ratings of the various training modes were only statistically significant for Statistics. For this skill, the results were consistent with the professional skills. Respondents liked learning Statistics through a university course, but not through self-taught or peer-taught mechanisms. In fact, participants' rejections of learning statistics through these mechanisms were two of the most statistically powerful results in the entire dataset, as seen in Table~\ref{tab:pvalues}.

\subsubsection{Mathematical Skills}
\label{sec:mathskills}

\begin{figure}
{\includegraphics[width=3.4in]{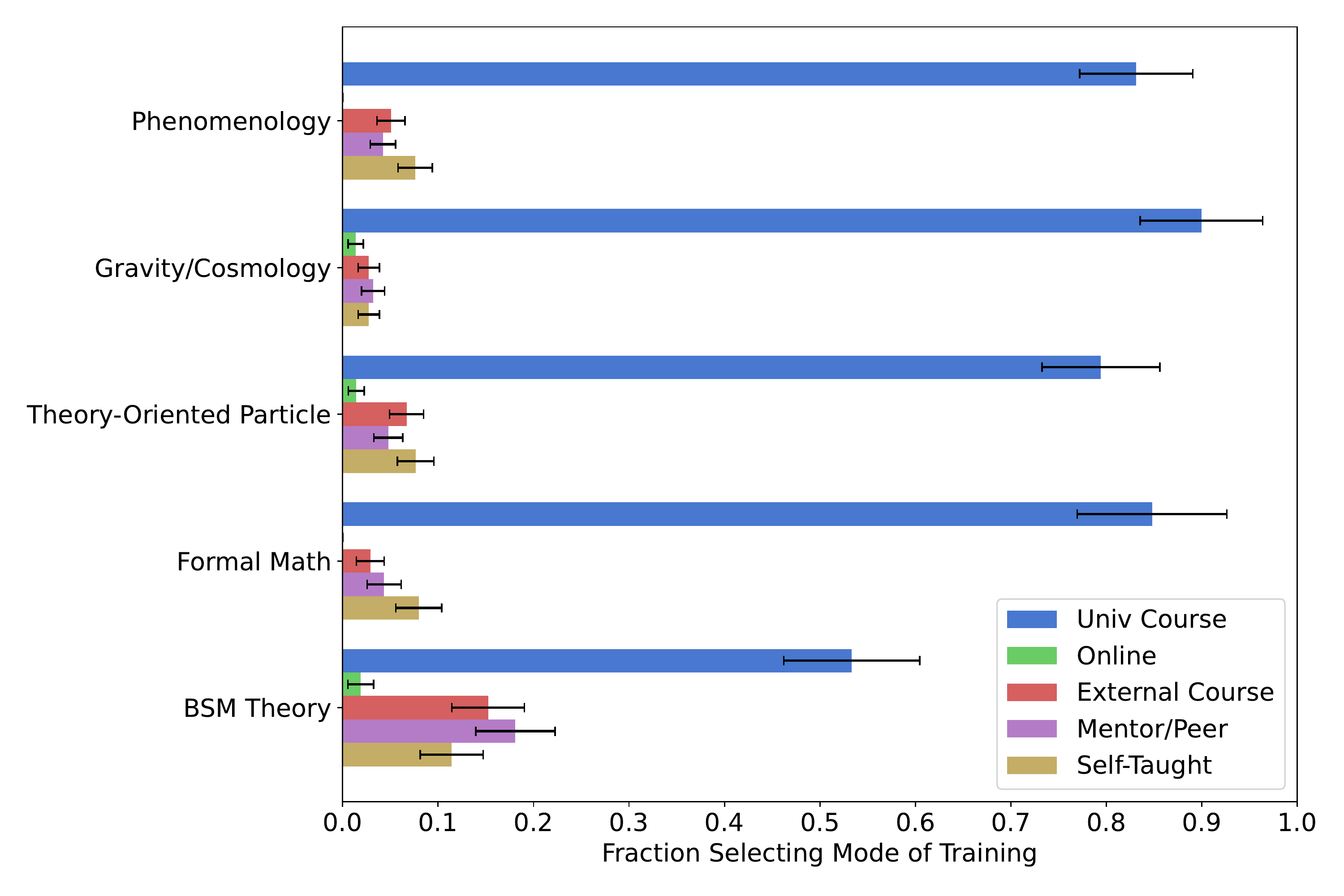}}
\caption{For each skill, we display the fraction of respondents selecting each mode out of all respondents who chose a mode. This question was not mandatory, so a few respondents said they gained experience in a skill, but did not select a modality; these responses are not included in the denominator. The five results for each skill add up to 1. Statistical Poisson errors are shown, calculated ($\sqrt{\text{N~``Mode''~responses}}$~/~N~``Any~Mode''~responses). ``Career-Level'' includes research staff, nontenured, and tenured faculty. The bars for each skill are stacked in the same vertical order as in the legend. The legend is abbreviated; see Section~\ref{sec:gradskills} for the exact wording of the modes.}
\label{fig:Mathematical_Mode}
\end{figure}

\begin{figure*}
{\includegraphics[width=6.8in]{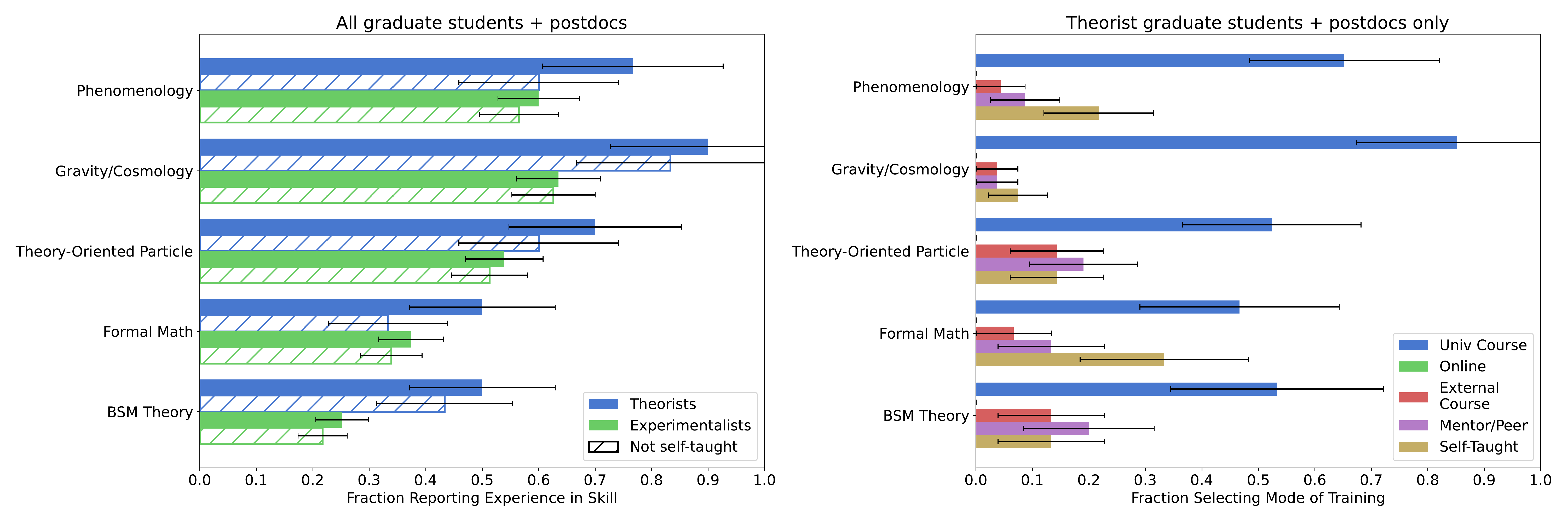}}
\caption{\textbf{(Left)} Comparison of theoretical vs.~experimental physicists' self-reported experience in mathematical skills. \textbf{(Right)} Only theoretical physicists' self-reported training modalities for these skills. Experimental physicists were almost exclusively taught these skills through university coursework, so these responses are skipped. \textbf{(Both)} Only graduate students and postdocs were asked about their research topic, so responses from other job categories are excluded.}
\label{fig:Mathematical_TheoryVsExpt}
\end{figure*}

\begin{figure}
{\includegraphics[width=3.4in]{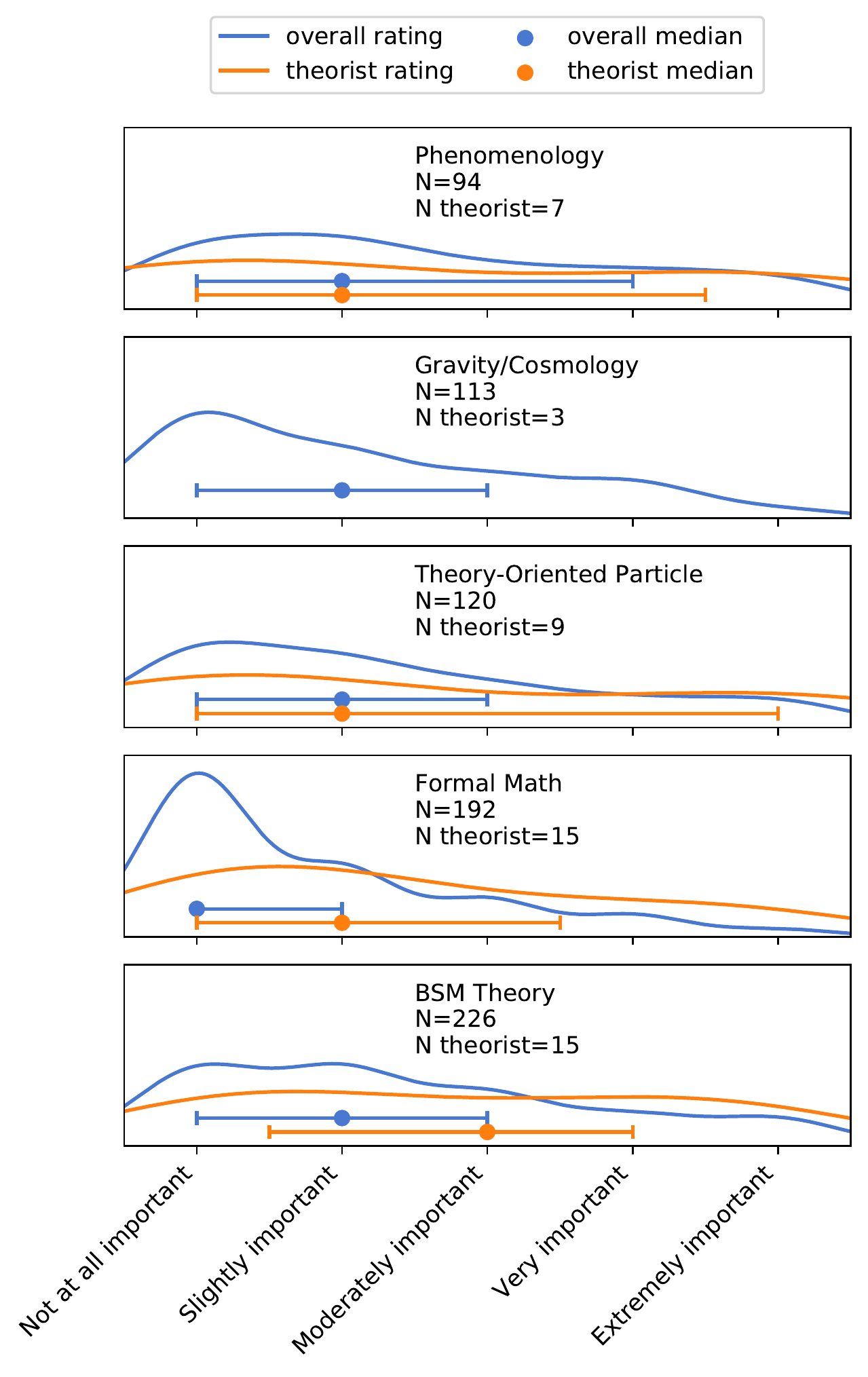}}
\caption{The distribution of respondents selecting each rating of importance to their career so far for each theoretical mathematical skill, from among those who did not receive the skill. Height is proportional to the number of responses with a given rating. The importance ratings from only those currently intending on a career in theory are shown separately in orange (excluding skills with N$<5$, i.e. Gravity/Cosmology). Sample sizes are quite limited due to the fact that theorists represent a small fraction of total responses and only those not receiving the listed skill were asked the corresponding question.}
\label{fig:Violin_Importance_Math}
\end{figure}

In the final skills section, we asked about a variety of mathematical skills, primarily geared towards theoretical physicists. Please see the survey text for the exact wording; most of these skills are abbreviated on plots. At first glance, the results are quite clear from Figs.~\ref{fig:Mathematical_Experience}~and~\ref{fig:Mathematical_Mode}. For the skills we asked about, there was variation in how many respondents gained these skills. Some skills, including phenomenology, gravity/cosmology, and theory-oriented particle physics, were obtained by a majority of respondents; others, including formal mathematics and Beyond the Standard Model physics, were gained by only a sizable minority. The vast majority of respondents gained these skills through university coursework.

Upon closer examination, a few interesting trends appear. From Table~\ref{tab:age_dependent_skills}, we see that for formal math, formal training is \textit{positively} correlated with career stage. This means that current early career physicists are more likely to have gained this skill through self-teaching and peer-teaching. Since this is the only skill with a positive correlation, we examined it more closely. Grad students and postdocs were about 10-15\% likely to gain formal math skills through self-teaching; for career-level physicists, this number was less than 5\%. We also see from Table~\ref{tab:pvalues} that, similar to the professional and technical skills, respondents prefer to gain mathematical skills through university coursework rather than through self-taught or peer-taught mechanisms.

One interesting result emerges if we separate our dataset into theorists and experimentalists, as done in Figure~\ref{fig:Mathematical_TheoryVsExpt}. Unfortunately, the question of research field was only asked to graduate students and postdocs, so they are the only respondents in this subset. Regardless, it's clear that theorists were more likely to experience these skills at all and learn them through informal modes. Very few experimentalists gained these skills by teaching themselves, but a fair amount of theorists did. A noticeable number of theorists also gained mathematical skills through peer- or mentor-led training. Recall that respondents were dissatisfied with both these modes, indicating a clear gap in theoretical particle physics education.

Overall, the theoretical mathematical skills asked about received somewhat low importance ratings for respondents' careers (Figure~\ref{fig:Violin_Importance_Math}), with perhaps slightly higher ratings for those skills respondents were most likely to be trained in (Figure~\ref{fig:Mathematical_Experience}), as for other skill categories. Ratings from just those graduate or postdoc respondents currently intending on a career in theory are analyzed separately, and show marginally higher ratings, consistent with the expectation that theoretical mathematical skills are more important for theorists. However, they remain relatively low in importance overall, perhaps due in part to potential bias introduced by asking for importance ratings only among those who did not receive experience in those skills.

\subsection{Correlations between skills, satisfaction, and career preparedness} \label{sec:preparedness}

We would like to understand whether training during graduate school is related to career preparedness. To address this, a correlation analysis using Kendall's $\tau$ coefficient was performed between the presence or absence of training for specific skills (excluding responses in the self-taught category) and reported preparation for various career tracks. We find, in general, fairly weak but positive correlations between training and career preparation. The only moderately-strong correlations ($\tau\ge0.2$) observed were between training in phenomenology or BSM theory and with preparation for university positions ($\tau=0.21$ and 0.22, respectively). 

Because individual skills are likely to have significant variation, we also considered correlations between the total number of trained (not self-taught) skills in each of the three categories and career preparation. However, this analysis also yielded mostly weak correlations: the strongest correlation observed was $\tau=0.20$ between total professional skills and reported preparation for the respondent's intended career. Many of the technical experimental and theoretical math skills here received ratings of ``slightly important'' or ``not at all important`` to respondent's careers so far, in accordance with these relatively weak correlations, suggesting that other skills not mentioned in the survey may be more important. The stronger correlation between career preparation and professional skills is, similarly, in accordance with their higher importance ratings, though it and the correlation between career preparation and high-importance technical skills (such as computer programming) are somewhat smaller than expected.

Similarly, receiving training in various skills was only weakly correlated with reported satisfaction in graduate school or specifically in particle physics. The only moderately-strong correlations observed were between training in phenomenology or overall (summed) math skills and particle physics-specific satisfaction ($\tau=0.28$ and 0.22, respectively). We interpret this as a sign that, when considering overall satisfaction with graduate education, training and skills received are not the most salient features.

We find stronger correlations between preparation for career paths and satisfaction in graduate education, primarily for more traditional academic career paths. There is a strong correlation between satisfaction in education and preparation for both university career paths ($\tau=0.42$) and national lab paths ($\tau=0.34$), a moderate correlation with preparation for industry ($\tau=0.20$), and a weak correlation with preparation for teaching K-12 ($\tau=0.12$). The stronger correlation with university and national lab careers is likely in part a reflection of the fact that these are the desired careers of a significant portion of respondents (who are unlikely to be disappointed by an education that did not prepare them for careers they are uninterested in). However, the correlation of a respondent's satisfaction with their intended career path is somewhat weaker (though still moderate, $\tau=0.28$). This is explained by the fact that those intending to pursue non-academic careers are not particularly unsatisfied with their education, even if it did not prepare them for their intended careers ($\tau=0.09$ for those intending to enter industry).

We also find strong correlations between initial preparation to meet advisor expectations and both overall graduate satisfaction ($\tau=0.33$) as well as preparation for both university ($\tau=0.35$) and lab ( $\tau=0.33$) positions. While we cannot establish causality, this is highly suggestive that undergraduate training is an important factor in ensuring both satisfying graduate experiences and academic career prospects beyond graduate school.

\begin{table*}
    \begin{tabular}{p{3.0in} p{0.7in} p{3.0in} }
          \begin{tabular}[c]{@{}l@{}}\textbf{ \large Conclusions} \end{tabular} 
          &
          \begin{tabular}[c]{@{}l@{}}\end{tabular}
          &
          \begin{tabular}[c]{@{}l@{}}\textbf{ \large Improvements}\end{tabular} 
        \\ \hline\hline\\

         Communication barrier to reaching a large number of undergraduates. &   &Mass communication needed.  
         \\\\
        Strong biases towards research experiences with university professors at home institution.  &    &Reach out to REU and DOE lab interns for future studies. 
        \\\\
        Significant biases towards R1 research universities.  &    &Actively reach out to PUIs, Liberal Arts Colleges, and Community Colleges for future studies. 
        \\
        \\

\hline\hline
\end{tabular}
\caption{\label{tab:table4}\textbf{Left:} Conclusions from the results of the Undergraduate Survey. \textbf{Right:} Proposed improvements to gather enough statistics to perform a future, more robust analysis of an undergraduate survey.}
\end{table*}

\section{Undergraduate Education Results}
\label{sec:UG}

Only respondents who selected “Undergraduate
Student” in the first question, “What position do you currently hold?, ” were shown the Undergraduate portion of this survey. As shown in Table~\ref{tab:survey_demographics}, this meant that we had only 24 participants answering the questions pertaining solely to undergraduate particle physics education. This represented only 6.7 percent of those taking our survey, and thus we are unable to draw definitive conclusions from the undergraduate portion of the analysis, given the lack of statistics.

Although we cannot go into as much depth as the graduate portion of the analysis, there are still several important takeaways that should be noted for a future undergraduate-focused survey to address. The first and most obvious is that the HEPA community needs some sort of mass communication tool or another mechanism to ascertain the needs of students (particularly those interested in particle physics) at the undergraduate level. Secondly, our answers were heavily biased towards research experience only with a university professor, and we lacked any input from students in REU and/or DOE lab internships. Lastly, we had a significant bias towards R1 and heavily research-based universities, with very few student responses reflecting input from Liberal Arts, teaching-focused, and community colleges. These are all summarized in Table~\ref{tab:table4}.

In order to further discuss the rationale underlying these three important conclusions, the following sections have been included. 

\subsubsection{Communication Barrier}

We recognize that the purpose of the physics curriculum at the undergraduate level is more to deliver breadth than depth on physics topics, and to allow students to survey sub-fields of research and/or careers. This approach seeks to provide a bedrock for further exploration and specialization at the graduate level. However, college students do have opportunities to become engaged with research in particle physics well before matriculation into a graduate degree program. This may be through a summer internship at a DOE lab or a small project with a professor at their university, or by simply taking an elective offered by their physics department. As such, steps should be taken to ensure that undergraduates, especially those who have had some exposure to particle physics, have the opportunity to remain connected with that community. In part, this would help to stem the flow of the ``leaky pipeline'' problem, but it would also foster a continuity for students to remain a part of and integrated into HEPA as they grow and progress in their own physics educational journey \cite{hsu_2015, staff_sousa_2020}.

Upon further reflection of how we chose to advertise this survey, we recognize that reaching many undergraduates was difficult, particularly those not connected through Snowmass or to an active particle physics working group and/or experiment. Despite our best efforts, we only managed to reach 24 undergraduates with HEPA connections in our survey. However, as stated above, media do exist to give students particle physics exposure during undergraduate studies (albeit, these media can and should be improved). Therefore, there must be a communication barrier to reach undergraduates, as the mechanisms we used did reach enough graduate students to provide a statistically significant sample.

One way to address this discrepancy would be to create and support a mass communication platform through which undergraduate students can more easily be contacted. We note that the DOE does provide internship participation information, but communication to said students remains private. One example is from the Science Undergraduate Laboratory internship (SULI) program in the DOE Office of Science \cite{science_2022}. University sponsored REUs may have the same information. However, a centralised way to contact students, especially those who did a particle physics project as an undergraduate, is crucial. This would provide not only an increased number of survey participants, but would allow future investigators to solicit input from a large subset of undergraduates who had a chance to see how their physics education (thus far) has impacted their particle physics research experience and future career goals and aspirations. We would therefore recommend, at minimum, that the DOE Office of Science (which oversees the National Lab internships) and REU programs at universities compile a list of contact information for recent participants in their programs; this could be made available to investigators conducting surveys of undergraduate particle physics education and research experience. Due to privacy concerns, we recommend that a central contact be made for each program, so a survey administrator can facilitate communication to these students.

We also note that our survey missed students who don't have research experience but might have had classwork and/or conference experiences in particle physics. Since we had several questions focusing on fundamental physics education as well as professional development topics, it is important for all undergraduates to take part, not merely those already doing research. One way to address this in a future survey is to have a central mass communication platform for physics departments across a variety of college or university types. 

In retrospect, we believe that it would have been beneficial to reach out directly to physics departments at R1 universities, Liberal Arts schools, Community Colleges, and other Primarily Undergraduate Institutions (PUI). We chose not to do so in this survey because we were concerned about inducing extra biases from our selection of who we reached out to. Although communicating with professors who are engaged in HEPA research and/or Snowmass is a good first step, manually contacting university physics departments to reach more undergraduates is not ideal. Therefore, there should be a system to easily and effectively contact physics departments across the country in order to better ascertain the needs of undergraduates, particularly those who have had any exposure---research or otherwise---to the field of particle physics.

\subsubsection{Consideration of REUs and DOE Internships}

While the particle physics community should consider expanding and strengthening the undergrad pipeline to sustain a prolific future in HEPA, it still remains difficult to reach a large number of undergraduate students who are interested and/or researching in particle physics. In large part, this is due to the fact that while there exists a plethora of undergraduate REU programs and internships around the country, the HEPA community does not have a networking apparatus for students to stay engaged and connected to the particle physics community before matriculating into a graduate program. If a student is able to find a research advisor at their institution who is willing to mentor them during their remaining time as an undergraduate, then it is possible for the student to remain engaged and connected with the HEPA community. However, when students are interns at labs for a summer or two and return to their home-institution, it can be difficult, if not impossible, to track their continued progress and interest in particle physics. This makes it difficult to contact and survey students who have had undergraduate particle physics experience. 

Another consideration is that students, even with a host adviser at their institution, should be recognized for their full collaborative capacity. In other words, they should be acknowledged, whether through meeting/conference presentations or in some other formal ways (at the discretion of a collaboration in the case of experiment groups) that highlight and recognize their contributions and efforts to further scientific advancement.

Therefore, we also call on HEPA collaborations to evaluate how they integrate all undergraduate researchers, DOE/REU students or otherwise, fully into their communities. We also strongly urge researchers working with undergraduate students to encourage them to join the collaboration as grad students and/or connect them with other collaborations.

Considering how important REU experiences and lab internships are in progression to graduate school PhD programs, we are missing a significant group of students who already have taken advantage of research opportunities. This should be rectified. We would like to recommend having a networking platform supported by the participating universities, DOE, and APS to facilitate a continued inclusion of students as they progress on their educational journey \cite{santos_bitter_kravitz_2020}. Not only would this help solve the communication barrier issue addressed in the previous section; it would also serve as a means to support students as they decide on future careers and provide resources to remain a part of the HEPA community. 

While APS offers undergraduate conferences such as CUWiP (Conference for Undergraduate Women in Physics), they could and should extend conferences for undergraduates to promote and sustain involvement and networking \cite{American_Physical_Society}. We look to efforts from the HEPA community to facilitate a particle physics conference for students, particularly undergraduates. One step in this direction would be to enlist the help of HEPA-focused universities and DOE labs---in particular, student groups at these institutions that are already organizing conferences designed for increased opportunities in HEPA networking. The annual New Perspectives Conference held at Fermilab is one successful example \cite{FSPA}. Undergraduates have been known to participate in this event, but there needs to be better communication and support to bolster this and other networking opportunities within HEPA. For instance, students from REUs and DOE internships should be encouraged to present their work at such a venue and to continue to participate each year, whether actively researching or not. 

Note that the Division of Nuclear Physics already sponsors Undergraduate-focused Conferences \cite{uw}. This is another example of an existing program that services the needs we recommend, and which can be used as a model to start a similar conference within the particle physics sector. 

\subsubsection{Inclusion of Teaching-based Universities and Colleges}

Some schools do not have particle physics classes at the undergrad level, particularly those with small undergraduate physics cohorts. Certainly, this is where it would be crucial to have research opportunities such as REUs or internships easily attainable for interested students to explore this field. However, regardless of offering particle physics research and/or classwork experiences, it is important to include schools that are primarily teaching-focused (as opposed to research-focused) in an analysis of undergraduate particle physics education. This is something that was lacking from the respondent turnout in our survey. 

While the primary purpose of our survey was to ascertain needs and skills for success in particle physics, it is also valuable to gauge the overall academic and professional opportunities that undergraduate students received at their institutions. Fundamental core classes and experiences which form the backbone of a typical undergraduate physics curriculum are instrumental in preparation for graduate school success and for particle physics research more broadly. Thus, we would be remiss if we didn't stress the need to consider the general physics preparation at the undergraduate level. We therefore recommend that future attempts to survey undergraduates actively reach out to departments not only at R1 universities but also PUIs, including Liberal Arts Colleges and Teaching Universities. 

To reach these institutions, we must repeat our first major takeaway that communication is key. We and other investigators should remember to contact departments from a variety of institutions; note that by implementing a mass communication system, investigators would reach many different types of schools without introducing severe selection bias. This would be especially crucial to obtain a robust analysis that includes undergraduates with and without particle physics classwork and/or research experience.

Finally, we note that community colleges should be included along with teaching-focused schools. While it is unlikely that students there have had a class or research experience in particle physics, including their input is important for the analysis. There are some community college external research experience opportunities, such as DOE CCI internships, but there could and should be more \cite{scienceCC}. Including community colleges in a future analysis would help us better understand the varied experiences of undergraduate students, especially from introductory physics classes. These introductory courses are  often smaller at community colleges than R1 universities, and they provide a unique approach to learning. This is true of many PUIs as well. In fact, in recent years, there has been a push for more inclusion of community colleges in HEPA research efforts. This push provides an important opportunity for students to experience particle physics research for the first time \cite{bellis_bhattacharya_hogan_malik_pearson_demuth_laureto_2022}. Regretfully, our survey only received one response from a community college student (Table~\ref{tab:table4}), emphasizing the need for better communication with this group.


\section{Discussion}
\label{sec:discussion}

Our survey shows a variety of conclusions about the state of undergraduate and graduate particle physics education in the US. At a high level, we observe significant gaps in career preparation, skills training, expectation-setting for graduate programs, and inclusion of undergraduates into the community. We also see hints of the strategies that can fill those gaps and areas for further investigation.

From the analyses presented in this paper, we are able to synthesize a collection of recommendations. These takeaways must be implemented by the particle physics community as a whole, but there are specific members who have the power to make change. These include several institutions: university departments, national lab directors and divisional heads, professional societies (APS most prominently, but also AIP, AAS, and others), and funding agencies such as NSF and DOE. In some of our recommendations, we suggest who the implementing body should be, but where not explicitly stated, we leave this question for future discussion.

\subsection{Recommendations for Graduate Education}\label{sec:actiongrad}

\begin{enumerate}

    \item Particle physics education must expand to consider non-HEPA career trajectories and prepare students for careers in industry, teaching, and other tracks. Scholars who have pursued or are successfully pursuing careers in HEPA feel like they were prepared for HEPA, but not necessarily for other careers. Much of this work should be done by university departments. Graduate programs in particle physics should normalize training for industry positions via encouragement of industry partnerships (such as summer research internships) and formal development of skills in demand beyond academia (such as computer programming, team management, and effective communication). Professional societies, meanwhile, can build connections between academics and industry professionals and host training programs for these skills. We also see a role for funding agencies to financially support these trainings.
    
    \item Examining our results through the lens of \cite{irvingBecomingPhysicistRoles2015}, the particle community should define the expected mathematical, technical, and professional skills that are necessary for a scholar to succeed in HEPA. The community should also work with professionals in other career tracks to identify the necessary skills in those careers. Professional societies are best-suited to collect these findings, which should be publicly available for current and prospective students.
        
    \item Universities and physics departments should provide undergraduate students with a more complete picture of what particle physicists do beyond classroom discussion of physics theory, such as increased opportunities for learning about research (e.g. seminars). They should also provide a realistic view of common career paths post-PhD in particle physics, including the breakdown of theory and experimental academic positions and the commonality of shifting to a non-academic career. For example, undergraduate programs could hold job workshops or host panels with scientists at different career stages. Implementation of this recommendation will help students make a more informed decision about what to study in graduate school and whether such a choice aligns with their goals. It will also allow students to be more prepared for graduate school, leading to greater satisfaction.
    
    \item Graduate programs in particle physics should support more formal modes of training for those skills where self-teaching is inadequate, indicated in Table~\ref{tab:pvalues}. As many of the professional skills are equally useful throughout physics disciplines or even other scientific and non-scientific fields, this could take many forms such as university-wide workshops, one-on-one coaching sessions, or shared online resources. It is critical for advisors or program coordinators to make their students aware of such resources and to actively encourage their use as part of their graduate training (and not a ``free time'' activity). In addition to university coursework, professional societies can also support external training programs in these skills, which respondents rated highly when there was sufficient data.
    
    \item Physics departments should consider making a statistics course a mandatory part of the undergraduate physics curriculum or strongly encourage it for particle physics-oriented students. Graduate programs could include formal statistics training as part of an upper-level topical course, e.g. mathematical methods, practical computing techniques, and statistics for particle physics.
    
    \item More formal training opportunities should be made available for advanced theoretical skills. These should primarily be done within a university environment, but to account for cases where a dedicated course may be too specialized or a suitable teacher not available, the community should provide opportunities such as virtual workshops which are free to attend. These should be supported by funding agencies.

\end{enumerate}

\subsection{Recommendations for Undergraduate Education}\label{sec:actionundergrad} 

\begin{enumerate}
    \item Actively plan to perform this survey again in the future for undergraduate data once the communication barriers listed in other recommendations are rectified. 
    
    \item Develop connections with those physics students who have elected industry-affiliated careers, perhaps through department alumni lists. This population is very important to survey -- we could tell more about the necessary skills for all future careers. 
    
    \item Create and support a mass communication platform by which undergraduate students can more easily be contacted. This would help to reach more college students so as to have enough statistics to accurately ascertain the needs of the undergraduate physics student population in a later survey. See the following two other notes for further details. This should be supported by the DOE Office of Science (who oversees the National Lab internships). Individual REUs at universities should release or compile a list of contact information for recent undergrad students in their respective programs. In that way we can ensure that those students who we know have had research experiences in HEPA are contacted and included in the future. 
    
    \item Create a central mass communication system for physics departments across a variety of college or university types - both R1 universities and Liberal Arts, Community Colleges, and PUIs. This will ensure input from teaching-focused institutions. 
    
    \item Provide a networking platform supported by universities, DOE, and APS for students to continue to stay engaged and connected to the particle physics community before matriculation into a graduate program. This is especially important for students from REUs and/or internships that do not have HEPA research options at their home institution. 
    
    \item Support HEPA universities and DOE labs that have student groups already organizing conferences to include and actively help in providing increased opportunities for undergraduates and graduate students to network with in HEPA community.
    
    \item  Call on HEPA collaborations to evaluate the integration of all researchers into their communities. That is to say HEPA collaborations should be required to confirm that all researchers --- including undergraduates, faculty/researchers at PUI and teaching focused institutions, engineers and technicians~\cite{hansenClimateFieldSnowmass2022} --- engaging in HEPA research are acknowledged, whether through publications and/or as official collaborators (in the case of experiments) for their contributions and efforts to further scientific advancement, no matter how small.

\end{enumerate}

\subsection{Recommendations for Future Surveys}\label{sec:missingquestions}  

As this study was developed by physics students and postdoctoral researchers, there are topics that we missed and changes we would make on hindsight. We list them here for use in the suggested future studies.

\begin{itemize}
    \item Section \ref{sec:gradcareers}: Survey the changes in career goal as a function of all positions. We did not ask this of career-level participants; had we done so, we would have been able to derive the flow of HEPA community members from the start of grad school, through commencement, early career choices, and final career category. 
    
    \item Section \ref{sec:subfieldprefs}: Survey the changes in subfield as a function of all positions. We would have liked to ask tenured/tenure track faculty and research staff if they had changed subfields during their career. 
    
    \item Section \ref{sec:subfieldprefs}: Survey career-level current subfields, so that all positions could be compared to aquisition of skills in graduate school, with subfield-category breakdown. 
    
    \item Section \ref{sec:gradskills}: Survey the importance of skills regardless of whether they were experienced during graduate study or not.
\end{itemize}


\section{Conclusion}
\label{sec:conclusion}

As part of the 2021 Snowmass Community Planning Process, we developed a survey to study mismatches between the skills necessary for careers in particle physics and the methodology of instruction for each of those skills. We collected 357 responses: a combination of undergraduate and graduate students, postdocs, faculty, research staff, and technicians. Respondents perceived themselves to be well-prepared for academic or laboratory careers, but training for industrial careers was found to be lacking. This is an issue of concern, given the large number of physicists who pursue careers outside academia. We also find that several skills seen as important for a HEPA career are being taught informally, when respondents prefer these skills to be taught through university-led or external courses. We recommend that the particle physics community evaluate the methods by which our members are developing these skills and create new pathways for junior physicists. In addition, the community should take steps to be more inclusive of undergrads, who are often involved in particle physics research but do not have mechanisms to stay engaged beyond the length of an individual project. The United States is a global leader in particle physics research, but our educational processes leave substantial room for improvement. If the entire community comes together to address the issues raised here, our early-career scientists will be more prepared for careers within and outside physics.

\section*{Acknowledgements}
The authors would like to acknowledge Yue Wang (University of California, Berkeley) for their input on section \ref{sec:gradresults} (graduate perspectives and interpretations) as well as their advice on on communication of these data through figures and tables, and Andrew Santos (\'{E}cole Polytechnique, Leprince-Ringuet Laboratory) for their invaluable discussions, thoughts, and comments on section \ref{sec:UG} as well as their contributions to an Undergraduate-focused LOI that influenced much of the content in section \ref{sec:UG}. Thanks also to Brenda Belcher (Office for Protection of Human Subjects, University of California, Berkeley) for their guidance in development and distribution of the survey, and to the Department of Physics, University of California, Berkeley, for their support of this postdoctoral-lead research project. Thanks also to the Snowmass leadership, especially the leadership of the Community Engagement Frontier and the Physics Education Topical Group, for useful discussions and their assistance in publicizing the survey.

We also thank the authors of the following letters of intent, submitted to the 2021 Snowmass Community Planning Exercise for their inspiration: 
\begin{itemize}
    \item A. Santos, O. Bitter, and S. Kravitz, ``Understanding and maximizing access to particle physics at the undergraduate level" \cite{santos_bitter_kravitz_2020}
    \item V. Velan \textit{et.al.} ``Enhancing Particle Physics Education at the Graduate Level" \cite{vetri_grad_LOI}
\end{itemize}

\clearpage
\setlength{\bibsep}{4pt}
\bibliography{Bibliography} 

\begin{thebibliography}{32}%
\makeatletter
\providecommand \@ifxundefined [1]{%
 \@ifx{#1\undefined}
}%
\providecommand \@ifnum [1]{%
 \ifnum #1\expandafter \@firstoftwo
 \else \expandafter \@secondoftwo
 \fi
}%
\providecommand \@ifx [1]{%
 \ifx #1\expandafter \@firstoftwo
 \else \expandafter \@secondoftwo
 \fi
}%
\providecommand \natexlab [1]{#1}%
\providecommand \enquote  [1]{``#1''}%
\providecommand \bibnamefont  [1]{#1}%
\providecommand \bibfnamefont [1]{#1}%
\providecommand \citenamefont [1]{#1}%
\providecommand \href@noop [0]{\@secondoftwo}%
\providecommand \href [0]{\begingroup \@sanitize@url \@href}%
\providecommand \@href[1]{\@@startlink{#1}\@@href}%
\providecommand \@@href[1]{\endgroup#1\@@endlink}%
\providecommand \@sanitize@url [0]{\catcode `\\12\catcode `\$12\catcode
  `\&12\catcode `\#12\catcode `\^12\catcode `\_12\catcode `\%12\relax}%
\providecommand \@@startlink[1]{}%
\providecommand \@@endlink[0]{}%
\providecommand \url  [0]{\begingroup\@sanitize@url \@url }%
\providecommand \@url [1]{\endgroup\@href {#1}{\urlprefix }}%
\providecommand \urlprefix  [0]{URL }%
\providecommand \Eprint [0]{\href }%
\providecommand \doibase [0]{https://doi.org/}%
\providecommand \selectlanguage [0]{\@gobble}%
\providecommand \bibinfo  [0]{\@secondoftwo}%
\providecommand \bibfield  [0]{\@secondoftwo}%
\providecommand \translation [1]{[#1]}%
\providecommand \BibitemOpen [0]{}%
\providecommand \bibitemStop [0]{}%
\providecommand \bibitemNoStop [0]{.\EOS\space}%
\providecommand \EOS [0]{\spacefactor3000\relax}%
\providecommand \BibitemShut  [1]{\csname bibitem#1\endcsname}%
\let\auto@bib@innerbib\@empty
\bibitem [{\citenamefont {Mulvey}\ \emph {et~al.}(2021)\citenamefont {Mulvey},
  \citenamefont {Nicholson},\ and\ \citenamefont
  {Pold}}]{mulveyTrendsPhysicsPhDs2021}%
  \BibitemOpen
  \bibfield  {author} {\bibinfo {author} {\bibfnamefont {P.}~\bibnamefont
  {Mulvey}, \bibfnamefont {J.}}, \bibinfo {author} {\bibfnamefont
  {S.}~\bibnamefont {Nicholson}},\ and\ \bibinfo {author} {\bibfnamefont
  {J.}~\bibnamefont {Pold}},\ }\href@noop {} {\emph {\bibinfo {title} {Trends
  in {{Physics PhDs}}}}},\ \bibinfo {type} {Tech. Rep.}\ (\bibinfo
  {institution} {{American Institute of Physics}},\ \bibinfo {year} {2021})\
  \bibinfo {note}
  {\url{https://www.aip.org/statistics/reports/trends-physics-phds-171819}}\BibitemShut
  {NoStop}%
\bibitem [{Emp(2020)}]{EmploymentCareersPhysics2020}%
  \BibitemOpen
  \href@noop {} {\emph {\bibinfo {title} {Employment and {{Careers}} in
  {{Physics}}}}},\ \bibinfo {type} {Tech. Rep.}\ (\bibinfo  {institution}
  {{American Institute of Physics}},\ \bibinfo {year} {2020})\ \bibinfo {note}
  {\url{https://www.aip.org/statistics/resources/initial-employment-physics-bachelors-and-phds-classes-2019-and-2020}}\BibitemShut
  {NoStop}%
\bibitem [{\citenamefont {Mulvey}\ and\ \citenamefont
  {Pold}(2020)}]{mulveyPhysicsDoctoratesSkills}%
  \BibitemOpen
  \bibfield  {author} {\bibinfo {author} {\bibfnamefont {P.}~\bibnamefont
  {Mulvey}}\ and\ \bibinfo {author} {\bibfnamefont {J.}~\bibnamefont {Pold}},\
  }\href@noop {} {\emph {\bibinfo {title} {Physics {{Doctorates}}: {{Skills
  Used}} and {{Satisfaction}} with {{Employment}} | {{American Institute}} of
  {{Physics}}}}},\ \bibinfo {type} {Tech. Rep.}\ (\bibinfo  {institution} {{AIP
  Statistical Research Center}},\ \bibinfo {year} {2020})\ \bibinfo {note}
  {\url{https://www.aip.org/statistics/reports/physics-doctorates-skills-used-satisfaction-employment-1516}}\BibitemShut
  {NoStop}%
\bibitem [{\citenamefont {Irving}\ and\ \citenamefont
  {Sayre}(2015)}]{irvingBecomingPhysicistRoles2015}%
  \BibitemOpen
  \bibfield  {author} {\bibinfo {author} {\bibfnamefont {P.~W.}\ \bibnamefont
  {Irving}}\ and\ \bibinfo {author} {\bibfnamefont {E.~C.}\ \bibnamefont
  {Sayre}},\ }\bibfield  {title} {\bibinfo {title} {Becoming a physicist:
  {{The}} roles of research, mindsets, and milestones in upper-division student
  perceptions},\ }\href {https://doi.org/10.1103/PhysRevSTPER.11.020120}
  {\bibfield  {journal} {\bibinfo  {journal} {Physical Review Special Topics -
  Physics Education Research}\ }\textbf {\bibinfo {volume} {11}},\ \bibinfo
  {pages} {020120} (\bibinfo {year} {2015})}\BibitemShut {NoStop}%
\bibitem [{\citenamefont {H{\"a}ussler}\ and\ \citenamefont
  {Hoffmann}(2002)}]{hausslerInterventionStudyEnhance2002}%
  \BibitemOpen
  \bibfield  {author} {\bibinfo {author} {\bibfnamefont {P.}~\bibnamefont
  {H{\"a}ussler}}\ and\ \bibinfo {author} {\bibfnamefont {L.}~\bibnamefont
  {Hoffmann}},\ }\bibfield  {title} {\bibinfo {title} {An intervention study to
  enhance girls' interest, self-concept, and achievement in physics classes:
  {{INTERVENTION STUDY}}},\ }\href {https://doi.org/10.1002/tea.10048}
  {\bibfield  {journal} {\bibinfo  {journal} {Journal of Research in Science
  Teaching}\ }\textbf {\bibinfo {volume} {39}},\ \bibinfo {pages} {870}
  (\bibinfo {year} {2002})},\ \bibinfo {note}
  {https://onlinelibrary.wiley.com/doi/10.1002/tea.10048}\BibitemShut {NoStop}%
\bibitem [{\citenamefont {Hazari}\ \emph {et~al.}(2013)\citenamefont {Hazari},
  \citenamefont {Sadler},\ and\ \citenamefont {Sonnert}}]{Hazari2013TheSI}%
  \BibitemOpen
  \bibfield  {author} {\bibinfo {author} {\bibfnamefont {Z.}~\bibnamefont
  {Hazari}}, \bibinfo {author} {\bibfnamefont {P.~M.}\ \bibnamefont {Sadler}},\
  and\ \bibinfo {author} {\bibfnamefont {G.}~\bibnamefont {Sonnert}},\
  }\bibfield  {title} {\bibinfo {title} {{The Science Identity of College
  Students: Exploring the Intersection of Gender, Race, and Ethnicity}},\
  }\href@noop {} {\bibfield  {journal} {\bibinfo  {journal} {{The Journal of
  College Science Teaching}}\ }\textbf {\bibinfo {volume} {42}},\ \bibinfo
  {pages} {82} (\bibinfo {year} {2013})}\BibitemShut {NoStop}%
\bibitem [{\citenamefont {Porter}(2019)}]{porterPhysicsPhDsTen2019}%
  \BibitemOpen
  \bibfield  {author} {\bibinfo {author} {\bibfnamefont {A.~M.}\ \bibnamefont
  {Porter}},\ }\href@noop {} {\emph {\bibinfo {title} {Physics {{PhDs Ten Years
  Later}}: {{Success Factors}} and {{Barriers}} in {{Career Paths}}}}},\
  \bibinfo {type} {Tech. Rep.}\ (\bibinfo  {institution} {{American Institute
  of Physics}},\ \bibinfo {year} {2019})\ \bibinfo {note}
  {\url{https://www.aip.org/statistics/reports/physics-phds-ten-years-later-success-factors-and-barriers-career-paths}}\BibitemShut
  {NoStop}%
\bibitem [{\citenamefont {Grodzins}(1971)}]{grodzinsManpowerCrisisPhysics1971}%
  \BibitemOpen
  \bibfield  {author} {\bibinfo {author} {\bibfnamefont {L.}~\bibnamefont
  {Grodzins}},\ }\href@noop {} {\emph {\bibinfo {title} {The {{Manpower
  Crisis}} in {{Physics}}}}},\ \bibinfo {type} {Tech. Rep.}\ (\bibinfo
  {institution} {{American Physical Society}},\ \bibinfo {year}
  {1971})\BibitemShut {NoStop}%
\bibitem [{\citenamefont {Malik}\ \emph {et~al.}(2022)\citenamefont {Malik},
  \citenamefont {Karadzhinova-Ferrer}, \citenamefont {Hogan}, \citenamefont
  {Bray}, \citenamefont {Kamalieddin}, \citenamefont {Flood}, \citenamefont
  {El-Zant}, \citenamefont {Fidalgo}, \citenamefont {Bruhwiler},\ and\
  \citenamefont {Bellis}}]{Snowmass_Careers_Whitepaper}%
  \BibitemOpen
  \bibfield  {author} {\bibinfo {author} {\bibfnamefont {S.}~\bibnamefont
  {Malik}}, \bibinfo {author} {\bibfnamefont {A.}~\bibnamefont
  {Karadzhinova-Ferrer}}, \bibinfo {author} {\bibfnamefont {J.}~\bibnamefont
  {Hogan}}, \bibinfo {author} {\bibfnamefont {R.}~\bibnamefont {Bray}},
  \bibinfo {author} {\bibfnamefont {R.}~\bibnamefont {Kamalieddin}}, \bibinfo
  {author} {\bibfnamefont {K.}~\bibnamefont {Flood}}, \bibinfo {author}
  {\bibfnamefont {A.}~\bibnamefont {El-Zant}}, \bibinfo {author} {\bibfnamefont
  {G.}~\bibnamefont {Fidalgo}}, \bibinfo {author} {\bibfnamefont
  {D.}~\bibnamefont {Bruhwiler}},\ and\ \bibinfo {author} {\bibfnamefont
  {M.}~\bibnamefont {Bellis}},\ }\href
  {https://doi.org/10.48550/ARXIV.2203.11665} {\bibinfo {title} {Facilitating
  non-hep career transition}} (\bibinfo {year} {2022})\BibitemShut {NoStop}%
\bibitem [{\citenamefont {Rethman}\ \emph {et~al.}(2021)\citenamefont {Rethman}
  \emph {et~al.}}]{rethmanImpactInformalPhysics2021}%
  \BibitemOpen
  \bibfield  {author} {\bibinfo {author} {\bibfnamefont {C.}~\bibnamefont
  {Rethman}} \emph {et~al.},\ }\bibfield  {title} {\bibinfo {title} {Impact of
  informal physics programs on university student development: {{Creating}} a
  physicist},\ }\href {https://doi.org/10.1103/PhysRevPhysEducRes.17.020110}
  {\bibfield  {journal} {\bibinfo  {journal} {Physical Review Physics Education
  Research}\ }\textbf {\bibinfo {volume} {17}},\ \bibinfo {pages} {020110}
  (\bibinfo {year} {2021})}\BibitemShut {NoStop}%
\bibitem [{\citenamefont {{ATLAS Experiment at
  CERN}}()}]{USAtlasCollaboration}%
  \BibitemOpen
  \bibfield  {author} {\bibinfo {author} {\bibnamefont {{ATLAS Experiment at
  CERN}}},\ }\href@noop {} {\bibinfo {title} {The {{Collaboration}}}},\
  \bibinfo {howpublished}
  {\url{https://atlas.cern/discover/collaboration}}\BibitemShut {NoStop}%
\bibitem [{\citenamefont {Good}\ \emph {et~al.}(2007)\citenamefont {Good},
  \citenamefont {Aronson},\ and\ \citenamefont
  {Harder}}]{goodProblemsPipelineStereotype2007}%
  \BibitemOpen
  \bibfield  {author} {\bibinfo {author} {\bibfnamefont {C.}~\bibnamefont
  {Good}}, \bibinfo {author} {\bibfnamefont {J.}~\bibnamefont {Aronson}},\ and\
  \bibinfo {author} {\bibfnamefont {J.~A.}\ \bibnamefont {Harder}},\ }\bibfield
   {title} {\bibinfo {title} {{Problems in the Pipeline: {{Stereotype}} Threat
  and Women's Achievement in High-Level Math Courses}},\ }\href
  {https://doi.org/10.1016/j.appdev.2007.10.004} {\bibfield  {journal}
  {\bibinfo  {journal} {{Journal of Applied Developmental Psychology}}\
  }\textbf {\bibinfo {volume} {29}},\ \bibinfo {pages} {17} (\bibinfo {year}
  {2007})},\ \bibinfo {note} {www.sciencedirect.com}\BibitemShut {NoStop}%
\bibitem [{\citenamefont {Beasley}\ and\ \citenamefont
  {Fischer}(2012)}]{beasleyWhyTheyLeave2012}%
  \BibitemOpen
  \bibfield  {author} {\bibinfo {author} {\bibfnamefont {M.~A.}\ \bibnamefont
  {Beasley}}\ and\ \bibinfo {author} {\bibfnamefont {M.~J.}\ \bibnamefont
  {Fischer}},\ }\bibfield  {title} {\bibinfo {title} {{Why They Leave: The
  Impact of Stereotype Threat on the Attrition of Women and Minorities from
  Science, Math and Engineering Majors}},\ }\href
  {https://doi.org/10.1007/s11218-012-9185-3} {\bibfield  {journal} {\bibinfo
  {journal} {Soc Psychol Educ}\ }\textbf {\bibinfo {volume} {15}},\ \bibinfo
  {pages} {427} (\bibinfo {year} {2012})}\BibitemShut {NoStop}%
\bibitem [{\citenamefont {Lewis}\ \emph {et~al.}(2016)\citenamefont {Lewis}
  \emph {et~al.}}]{lewisFittingOptingOut2016}%
  \BibitemOpen
  \bibfield  {author} {\bibinfo {author} {\bibfnamefont {K.~L.}\ \bibnamefont
  {Lewis}} \emph {et~al.},\ }\bibfield  {title} {\bibinfo {title} {{Fitting in
  or Opting out: {{A}} Review of Key Social-Psychological Factors Influencing a
  Sense of Belonging for Women in Physics}},\ }\bibfield  {journal} {\bibinfo
  {journal} {Physical Review Physics Education Research}\ }\textbf {\bibinfo
  {volume} {12}},\ \href {https://doi.org/10.1103/PhysRevPhysEducRes.12.020110}
  {10.1103/PhysRevPhysEducRes.12.020110} (\bibinfo {year} {2016})\BibitemShut
  {NoStop}%
\bibitem [{\citenamefont {Eddy}\ and\ \citenamefont
  {Brownell}(2016)}]{eddyNumbersReviewGender2016}%
  \BibitemOpen
  \bibfield  {author} {\bibinfo {author} {\bibfnamefont {S.~L.}\ \bibnamefont
  {Eddy}}\ and\ \bibinfo {author} {\bibfnamefont {S.~E.}\ \bibnamefont
  {Brownell}},\ }\bibfield  {title} {\bibinfo {title} {{Beneath the Numbers:
  {{A}} Review of Gender Disparities in Undergraduate Education across Science,
  Technology, Engineering, and Math Disciplines}},\ }\bibfield  {journal}
  {\bibinfo  {journal} {Physical Review Physics Education Research}\ }\textbf
  {\bibinfo {volume} {12}},\ \href
  {https://doi.org/10.1103/PhysRevPhysEducRes.12.020106}
  {10.1103/PhysRevPhysEducRes.12.020106} (\bibinfo {year} {2016})\BibitemShut
  {NoStop}%
\bibitem [{\citenamefont {Spencer}\ \emph {et~al.}(1999)\citenamefont
  {Spencer}, \citenamefont {Steele},\ and\ \citenamefont
  {Quinn}}]{spencerStereotypeThreatWomen1999}%
  \BibitemOpen
  \bibfield  {author} {\bibinfo {author} {\bibfnamefont {S.~J.}\ \bibnamefont
  {Spencer}}, \bibinfo {author} {\bibfnamefont {C.~M.}\ \bibnamefont
  {Steele}},\ and\ \bibinfo {author} {\bibfnamefont {D.~M.}\ \bibnamefont
  {Quinn}},\ }\bibfield  {title} {\bibinfo {title} {Stereotype {{Threat}} and
  {{Women}}'s {{Math Performance}}},\ }\href
  {https://doi.org/10.1006/jesp.1998.1373} {\bibfield  {journal} {\bibinfo
  {journal} {Journal of Experimental Social Psychology}\ }\textbf {\bibinfo
  {volume} {35}},\ \bibinfo {pages} {4} (\bibinfo {year} {1999})}\BibitemShut
  {NoStop}%
\bibitem [{\citenamefont {Carlone}\ and\ \citenamefont
  {Johnson}(2007)}]{carloneUnderstandingScienceExperiences2007}%
  \BibitemOpen
  \bibfield  {author} {\bibinfo {author} {\bibfnamefont {H.~B.}\ \bibnamefont
  {Carlone}}\ and\ \bibinfo {author} {\bibfnamefont {A.}~\bibnamefont
  {Johnson}},\ }\bibfield  {title} {\bibinfo {title} {Understanding the
  {{Science Experiences}} of {{Successful Women}} of {{Color}}: {{Science
  Identity}} as an {{Analytic Lens}}},\ }\href
  {https://doi.org/10.1002/tea.20237} {\bibfield  {journal} {\bibinfo
  {journal} {J Res Sci Teach}\ }\textbf {\bibinfo {volume} {44}},\ \bibinfo
  {pages} {1187} (\bibinfo {year} {2007})}\BibitemShut {NoStop}%
\bibitem [{\citenamefont {Cooper}\ and\ \citenamefont
  {Brownell}(2016)}]{cooperComingOutClass2016}%
  \BibitemOpen
  \bibfield  {author} {\bibinfo {author} {\bibfnamefont {K.~M.}\ \bibnamefont
  {Cooper}}\ and\ \bibinfo {author} {\bibfnamefont {S.~E.}\ \bibnamefont
  {Brownell}},\ }\bibfield  {title} {\bibinfo {title} {Coming {{Out}} in
  {{Class}}: {{Challenges}} and {{Benefits}} of {{Active Learning}} in a
  {{Biology Classroom}} for {{LGBTQIA Students}}},\ }\href
  {https://doi.org/10.1187/cbe.16-01-0074} {\bibfield  {journal} {\bibinfo
  {journal} {CBE Life Sciences Education}\ }\textbf {\bibinfo {volume} {15}},\
  \bibinfo {pages} {1} (\bibinfo {year} {2016})}\BibitemShut {NoStop}%
\bibitem [{\citenamefont {Jones}\ \emph {et~al.}(2013)\citenamefont {Jones},
  \citenamefont {Ruff},\ and\ \citenamefont
  {Paretti}}]{jonesImpactEngineeringIdentification2013}%
  \BibitemOpen
  \bibfield  {author} {\bibinfo {author} {\bibfnamefont {B.~D.}\ \bibnamefont
  {Jones}}, \bibinfo {author} {\bibfnamefont {C.}~\bibnamefont {Ruff}},\ and\
  \bibinfo {author} {\bibfnamefont {M.~C.}\ \bibnamefont {Paretti}},\
  }\bibfield  {title} {\bibinfo {title} {{The Impact of Engineering
  Identification and Stereotypes on Undergraduate Women's Achievement and
  Persistence in Engineering}},\ }\href
  {https://doi.org/10.1007/s11218-013-9222-x} {\bibfield  {journal} {\bibinfo
  {journal} {Soc Psychol Educ}\ }\textbf {\bibinfo {volume} {16}},\ \bibinfo
  {pages} {471} (\bibinfo {year} {2013})}\BibitemShut {NoStop}%
\bibitem [{\citenamefont {Kendall}(1970)}]{kendallCorrelation}%
  \BibitemOpen
  \bibfield  {author} {\bibinfo {author} {\bibfnamefont {M.}~\bibnamefont
  {Kendall}},\ }\href@noop {} {\emph {\bibinfo {title} {Rank correlation
  methods}}}\ (\bibinfo  {publisher} {Charles Griffin \& Co.},\ \bibinfo {year}
  {1970})\BibitemShut {NoStop}%
\bibitem [{\citenamefont {Agarwal}\ \emph {et~al.}(2022)\citenamefont {Agarwal}
  \emph {et~al.}}]{agarwalSnowmass2021Community2022}%
  \BibitemOpen
  \bibfield  {author} {\bibinfo {author} {\bibfnamefont {G.}~\bibnamefont
  {Agarwal}} \emph {et~al.},\ }\bibfield  {title} {\bibinfo {title} {Snowmass
  2021 {{Community Survey Report}}},\ }\href@noop {} {\  (\bibinfo {year}
  {2022})},\ \Eprint {https://arxiv.org/abs/2203.07328} {arXiv:2203.07328
  [astro-ph, physics:hep-ex, physics:hep-ph, physics:physics]} \BibitemShut
  {NoStop}%
\bibitem [{\citenamefont {Hsu}(2015)}]{hsu_2015}%
  \BibitemOpen
  \bibfield  {author} {\bibinfo {author} {\bibfnamefont {S.}~\bibnamefont
  {Hsu}},\ }\bibfield  {title} {\bibinfo {title} {{STEM, Gender, and Leaky
  Pipelines}},\ }\href@noop {} {\bibfield  {journal} {\bibinfo  {journal}
  {Spartan Ideas}\ } (\bibinfo {year} {2015})},\ \bibinfo {note}
  {\url{https://spartanideas.msu.edu/2015/02/20/stem-gender-and-leaky-pipelines/}}\BibitemShut
  {NoStop}%
\bibitem [{\citenamefont {Sousa}(2020)}]{staff_sousa_2020}%
  \BibitemOpen
  \bibfield  {author} {\bibinfo {author} {\bibfnamefont {M.}~\bibnamefont
  {Sousa}},\ }\bibfield  {title} {\bibinfo {title} {Weekender | {Fixing the
  Leaky Pipeline: Women in the Physical Sciences}},\ }\href@noop {} {\bibfield
  {journal} {\bibinfo  {journal} {The Daily Californian}\ } (\bibinfo {year}
  {2020})},\ \bibinfo {note}
  {\url{https://www.dailycal.org/2020/04/02/fixing-the-leaky-pipeline-women-in-the-physical-sciences/}}\BibitemShut
  {NoStop}%
\bibitem [{\citenamefont {{US Department of Energy, Office of
  Science}}(2022{\natexlab{a}})}]{science_2022}%
  \BibitemOpen
  \bibfield  {author} {\bibinfo {author} {\bibnamefont {{US Department of
  Energy, Office of Science}}},\ }\href@noop {} {\bibinfo {title} {Science
  undergraduate laboratory internship}} (\bibinfo {year}
  {2022}{\natexlab{a}}),\ \bibinfo {note}
  {\url{https://science.osti.gov/wdts/suli}}\BibitemShut {NoStop}%
\bibitem [{\citenamefont {Santos}\ \emph {et~al.}(2020)\citenamefont {Santos},
  \citenamefont {Bitter},\ and\ \citenamefont
  {Kravitz}}]{santos_bitter_kravitz_2020}%
  \BibitemOpen
  \bibfield  {author} {\bibinfo {author} {\bibfnamefont {A.}~\bibnamefont
  {Santos}}, \bibinfo {author} {\bibfnamefont {O.}~\bibnamefont {Bitter}},\
  and\ \bibinfo {author} {\bibfnamefont {S.}~\bibnamefont {Kravitz}},\
  }\href@noop {} {\bibinfo {title} {{Understanding and Maximizing Access to
  Particle Physics at the Undergraduate Level}}},\ \bibinfo {howpublished}
  {Snowmass2021 LOI} (\bibinfo {year} {2020}),\ \bibinfo {note} {\\
  \url{https://www.snowmass21.org/docs/files/summaries/CommF/SNOWMASS21-CommF4_CommF2-036.pdf}}\BibitemShut
  {NoStop}%
\bibitem [{\citenamefont {{US Department of Energy, Office of
  Science}}(2022{\natexlab{b}})}]{American_Physical_Society}%
  \BibitemOpen
  \bibfield  {author} {\bibinfo {author} {\bibnamefont {{US Department of
  Energy, Office of Science}}},\ }\href@noop {} {\bibinfo {title} {{APS
  Conferences for Undergraduate Women in Physics (CUWIP)}}} (\bibinfo {year}
  {2022}{\natexlab{b}}),\ \bibinfo {note}
  {\url{https://www.aps.org/programs/women/cuwip/index.cfm}}\BibitemShut
  {NoStop}%
\bibitem [{\citenamefont {{FSPA}}(2021)}]{FSPA}%
  \BibitemOpen
  \bibfield  {author} {\bibinfo {author} {\bibnamefont {{FSPA}}},\ }\href@noop
  {} {\bibinfo {title} {{Fermilab Student and Postdoc Association: New
  Perspectives 2021}}} (\bibinfo {year} {2021}),\ \bibinfo {note}
  {\url{https://fspa.fnal.gov/new-perspectives-2021/}}\BibitemShut {NoStop}%
\bibitem [{\citenamefont {{University of Wisconsin at LaCrosse}}()}]{uw}%
  \BibitemOpen
  \bibfield  {author} {\bibinfo {author} {\bibnamefont {{University of
  Wisconsin at LaCrosse}}},\ }\href@noop {} {\bibinfo {title} {{Conference
  experience for Undergraduates (CEU)}}},\ \bibinfo {note}
  {\url{https://www.uwlax.edu/ceu/}}\BibitemShut {NoStop}%
\bibitem [{\citenamefont {{US Department of Energy, Office of
  Science}}(2022{\natexlab{c}})}]{scienceCC}%
  \BibitemOpen
  \bibfield  {author} {\bibinfo {author} {\bibnamefont {{US Department of
  Energy, Office of Science}}},\ }\href@noop {} {\bibinfo {title} {Community
  college internships}} (\bibinfo {year} {2022}{\natexlab{c}}),\ \bibinfo
  {note} {\url{https://science.osti.gov/wdts/cci}}\BibitemShut {NoStop}%
\bibitem [{\citenamefont {Bellis}\ \emph {et~al.}(2022)\citenamefont {Bellis}
  \emph
  {et~al.}}]{bellis_bhattacharya_hogan_malik_pearson_demuth_laureto_2022}%
  \BibitemOpen
  \bibfield  {author} {\bibinfo {author} {\bibfnamefont {M.}~\bibnamefont
  {Bellis}} \emph {et~al.},\ }\bibfield  {title} {\bibinfo {title} {{Enhancing
  {{HEP}} Research in Predominantly Undergraduate Institutions and Community
  Colleges}},\ }\href@noop {} {\  (\bibinfo {year} {2022})},\ \Eprint
  {https://arxiv.org/abs/2203.11662} {arXiv:2203.11662 [hep-ex,
  physics:physics]} \BibitemShut {NoStop}%
\bibitem [{\citenamefont {Hansen}\ \emph {et~al.}(2022)\citenamefont {Hansen}
  \emph {et~al.}}]{hansenClimateFieldSnowmass2022}%
  \BibitemOpen
  \bibfield  {author} {\bibinfo {author} {\bibfnamefont {E.~V.}\ \bibnamefont
  {Hansen}} \emph {et~al.},\ }\bibfield  {title} {\bibinfo {title} {{Climate of
  the {{Field}}: {{Snowmass}} 2021}},\ }\href@noop {} {\  (\bibinfo {year}
  {2022})},\ \Eprint {https://arxiv.org/abs/2204.03713} {arXiv:2204.03713
  [physics.soc-ph]} \BibitemShut {NoStop}%
\bibitem [{\citenamefont {Velan}\ \emph {et~al.}(2020)\citenamefont {Velan}
  \emph {et~al.}}]{vetri_grad_LOI}%
  \BibitemOpen
  \bibfield  {author} {\bibinfo {author} {\bibfnamefont {V.}~\bibnamefont
  {Velan}} \emph {et~al.},\ }\href@noop {} {\bibinfo {title} {{Enhancing
  Particle Physics Education at the Graduate Level}}},\ \bibinfo {howpublished}
  {Snowmass2021 LOI} (\bibinfo {year} {2020}),\ \bibinfo {note} {\\
  \url{https://www.snowmass21.org/docs/files/summaries/CommF/SNOWMASS21-CommF4_CommF2-036.pdf}}\BibitemShut
  {NoStop}%
\end{thebibliography}%

\appendix
\section{Full Survey Text}
\label{app:full_survey}
\clearpage




\section*{Supplementary material}

**Transforming US Particle Physics Education: A Snowmass Study**

You are invited to participate in a research study designed to understand the current status in particle physics education in the United States, including any common deficiencies. We expect to evaluate opinions on curriculum & method of learning, based on experiences in education. We also expect to evaluate opinions on preparedness for future job performance.

The results of this survey will be compiled and published in the "Transforming US Particle Physics Education: A Snowmass Study" contributed paper, which will be publicly available on arXiv.org after submission. Results and recommendations from this survey may be included in the final report of the Snowmass project which will be provided to the Department of Energy as part of funding recommendations from the particle physics community.

Participation in this study is voluntary. There is no monetary compensation. The survey should take no more than 10 minutes.

Participation in this study may not benefit you directly, but it will help us provide recommendations for the evolution of particle physics education within the next 10 years.

All questions are voluntary; you may skip any question and continue the survey. The only exception is the first question, which allows us to tailor responses to certain members of the community.

The information you share with us will be de-identified immediately upon submission. This means any personally identifying information (including location and IP address) are *not retrievable by researchers, and never will be.*

If you have any questions about this study, please contact Erin V Hansen (evhansen@berkeley.edu) or Vetri Velan (vvelan@berkeley.edu). If you have questions about your rights as a research participant, you may contact the Berkeley Office For Protection of Human Subjects (https://cphs.berkeley.edu/) at ophs@berkeley.edu.

By continuing with this survey, you acknowledge that the following are true:

- I am currently located in the United States. (Due to EU GDP regulations, we cannot currently include participants who are located abroad.)
- I am over 18 years of age.
- I self-identify as a participant in the particle physics community.

Please print or save a copy of this decree for your records. You may also find a printable version of this declaration at [link to be added].

**By completing this survey, you are consenting to participate in this study.**

Collaborators on this study are also providing a resource for collaboration between particle physics educators. This forum is NOT part of the research study, and participation is entirely optional. No forum data will contribute to this research study. You may find the survey at the following link: [link to be added]

What position do you currently hold?

- ◯ Undergraduate Student
- ◯ Graduate Student
- ◯ Postdoctoral Researcher
- ◯ Tenure-track Faculty
- ◯ Tenured Faculty
- ◯ Research Staff (e.g. National Lab)
- ◯ Technician/Engineer (Academic)
- ◯ Technician/Engineer (Industry)

What types of institution(s) are you currently affiliated with? (Select all that apply)

- ☐ Public University, PhD-granting institution
- ☐ Private University, PhD-granting institution
- ☐ Liberal Arts College
- ☐ Community College
- ☐ For-Profit Institution
- ☐ National Laboratory
- ☐ [_____] Other

**GRAD Satisfaction**

How satisfied do you feel with your graduate education in physics?

Extremely
dissatisfied
◯

Somewhat
dissatisfied
◯

Neither satisfied
nor dissatisfied
◯

Somewhat
satisfied
◯

Extremely
satisfied
◯

How satisfied do you feel with your graduate education in *particle* physics?

Extremely
dissatisfied
○

Somewhat
dissatisfied
○

Neither satisfied
nor dissatisfied
○

Somewhat
satisfied
○

Extremely
satisfied
○

**Formal Instruction (Technical)**

**Technical Skills**

For each of these topics, please state if you received any experience in it during **your own** undergraduate + graduate education.

|  | Yes | No |
|---|---|---|
| Statistics | ○ | ○ |
| Computer programming | ○ | ○ |
| Machine learning | ○ | ○ |
| Electrical engineering | ○ | ○ |
| Mechanical engineering | ○ | ○ |
| Materials science | ○ | ○ |

For each of these topics that you did gain experience with during undergraduate + graduate school, what was the primary mode of training you received? (Select the last option "self-taught" if you received no training.)

| | In a formal course at your university | Online, through a structured program (such as edX or Coursera) | In a course organized outside of the classroom (e.g. summer school, CERN courses, professional workshop, formal training in your research group, etc.) | Mentoring or peer learning | As needed; in a decentralized manner, self-taught |
|---|---|---|---|---|---|
| » Statistics | ○ | ○ | ○ | ○ | ○ |
| » Computer programming | ○ | ○ | ○ | ○ | ○ |
| » Machine learning | ○ | ○ | ○ | ○ | ○ |
| » Electrical engineering | ○ | ○ | ○ | ○ | ○ |
| » Mechanical engineering | ○ | ○ | ○ | ○ | ○ |
| » Materials science | ○ | ○ | ○ | ○ | ○ |

Please rate the method of training you selected above.

| | Extremely dissatisfied | Somewhat dissatisfied | Neither satisfied nor dissatisfied | Somewhat satisfied | Extremely satisfied |
|---|---|---|---|---|---|
| » Statistics | ○ | ○ | ○ | ○ | ○ |
| » Computer programming | ○ | ○ | ○ | ○ | ○ |
| » Machine learning | ○ | ○ | ○ | ○ | ○ |
| » Electrical engineering | ○ | ○ | ○ | ○ | ○ |
| » Mechanical engineering | ○ | ○ | ○ | ○ | ○ |
| » Materials science | ○ | ○ | ○ | ○ | ○ |

For each of these topics that you did not experience during your undergraduate + graduate education, how important has this topic been in your career so far?

| | Not at all important | Slightly important | Moderately important | Very important | Extremely important |
|---|---|---|---|---|---|
| » Statistics | ○ | ○ | ○ | ○ | ○ |
| » Computer programming | ○ | ○ | ○ | ○ | ○ |
| » Machine learning | ○ | ○ | ○ | ○ | ○ |
| » Electrical engineering | ○ | ○ | ○ | ○ | ○ |
| » Mechanical engineering | ○ | ○ | ○ | ○ | ○ |
| » Materials science | ○ | ○ | ○ | ○ | ○ |

**Formal Instruction (Theory)**

**Mathematical / Physics Topics**

How well were you prepared overall to meet the mathematical side of your graduate physics courses?

Not well at all
○

Slightly well
○

Moderately well
○

Very well
○

Extremely well
○

For each of these topics, please state if you received any experience in it during **your own** undergraduate + graduate education:

|  | Yes | No |
|---|---|---|
| Phenomenology-oriented particle physics [e.g. group theory] | ◯ | ◯ |
| Theory-oriented particle physics [SUSY, Amplitudes, etc.] | ◯ | ◯ |
| Beyond the standard model theories [e.g. string theory, quantum gravity] | ◯ | ◯ |
| Mathematics for formal theories [Topology, Geometry, etc.] | ◯ | ◯ |
| Gravity, GR, and cosmology | ◯ | ◯ |

For each of these topics that you did gain experience with during undergraduate + graduate school, what was the primary mode of training you received? (Select the last option "self-taught" if you received no training.)

| | In a formal course at your university | Online, through a structured program (such as edX or Coursera) | In a course organized outside of the classroom (e.g. summer school, CERN courses, professional workshop, formal training in your research group, etc.) | Mentoring or peer learning | As needed; in a decentralized manner, self-taught |
|---|---|---|---|---|---|
| » Phenomenology-oriented particle physics [e.g. group theory] | ○ | ○ | ○ | ○ | ○ |
| » Theory-oriented particle physics [SUSY, Amplitudes, etc.] | ○ | ○ | ○ | ○ | ○ |
| » Beyond the standard model theories [e.g. string theory, quantum gravity] | ○ | ○ | ○ | ○ | ○ |
| » Mathematics for formal theories [Topology, Geometry, etc.] | ○ | ○ | ○ | ○ | ○ |
| » Gravity, GR, and cosmology | ○ | ○ | ○ | ○ | ○ |

Please rate the method of training you selected above.

| | Extremely dissatisfied | Somewhat dissatisfied | Neither satisfied nor dissatisfied | Somewhat satisfied | Extremely satisfied |
|---|---|---|---|---|---|
| » Phenomenology-oriented particle physics [e.g. group theory] | ○ | ○ | ○ | ○ | ○ |
| » Theory-oriented particle physics [SUSY, Amplitudes, etc.] | ○ | ○ | ○ | ○ | ○ |
| » Beyond the standard model theories [e.g. string theory, quantum gravity] | ○ | ○ | ○ | ○ | ○ |
| » Mathematics for formal theories [Topology, Geometry, etc.] | ○ | ○ | ○ | ○ | ○ |
| » Gravity, GR, and cosmology | ○ | ○ | ○ | ○ | ○ |

For each of these topics that you did not experience during your undergraduate + graduate education, how important has this topic been in your career so far?

| | Not at all important | Slightly important | Moderately important | Very important | Extremely important |
|---|---|---|---|---|---|
| » Phenomenology-oriented particle physics [e.g. group theory] | ○ | ○ | ○ | ○ | ○ |
| » Theory-oriented particle physics [SUSY, Amplitudes, etc.] | ○ | ○ | ○ | ○ | ○ |
| » Beyond the standard model theories [e.g. string theory, quantum gravity] | ○ | ○ | ○ | ○ | ○ |
| » Mathematics for formal theories [Topology, Geometry, etc.] | ○ | ○ | ○ | ○ | ○ |
| » Gravity, GR, and cosmology | ○ | ○ | ○ | ○ | ○ |

**Formal Instruction (Professional)**

**Professional skills**

For each of these topics, please state if you received any experience in it during **your own** undergraduate + graduate education:

| | Yes | No |
|---|---|---|
| Mentoring younger scientists | ○ | ○ |
| Teaching | ○ | ○ |
| Scientific writing | ○ | ○ |
| Giving presentations | ○ | ○ |
| Writing a grant application | ○ | ○ |
| Outreach to the public | ○ | ○ |
| Writing a CV | ○ | ○ |
| Job applications (academic) | ○ | ○ |
| Job applications (non-academic) | ○ | ○ |

For each of these topics that you did gain experience with during undergraduate + graduate school, what was the primary mode of training you received? (Select the last option "self-taught" if you received no training.)

| | In a formal course at your university | Online, through a structured program (such as edX or Coursera) | In a course organized outside of the classroom (e.g. summer school, CERN courses, professional workshop, formal training in your research group, etc.) | Mentoring or peer learning | As needed; in a decentralized manner, self-taught |
|---|---|---|---|---|---|
| » Mentoring younger scientists | O | O | O | O | O |
| » Teaching | O | O | O | O | O |
| » Scientific writing | O | O | O | O | O |
| » Giving presentations | O | O | O | O | O |
| » Writing a grant application | O | O | O | O | O |
| » Outreach to the public | O | O | O | O | O |
| » Writing a CV | O | O | O | O | O |
| » Job applications (academic) | O | O | O | O | O |
| » Job applications (non-academic) | O | O | O | O | O |

Please rate the method of training you selected above.

| | Extremely dissatisfied | Somewhat dissatisfied | Neither satisfied nor dissatisfied | Somewhat satisfied | Extremely satisfied |
|---|---|---|---|---|---|
| » Mentoring younger scientists | ○ | ○ | ○ | ○ | ○ |
| » Teaching | ○ | ○ | ○ | ○ | ○ |
| » Scientific writing | ○ | ○ | ○ | ○ | ○ |
| » Giving presentations | ○ | ○ | ○ | ○ | ○ |
| » Writing a grant application | ○ | ○ | ○ | ○ | ○ |
| » Outreach to the public | ○ | ○ | ○ | ○ | ○ |
| » Writing a CV | ○ | ○ | ○ | ○ | ○ |
| » Job applications (academic) | ○ | ○ | ○ | ○ | ○ |
| » Job applications (non-academic) | ○ | ○ | ○ | ○ | ○ |

For each of these topics that you did not experience during your undergraduate + graduate education, how important has this topic been in your career so far?

| | Not at all important | Slightly important | Moderately important | Very important | Extremely important |
|---|---|---|---|---|---|
| » Mentoring younger scientists | ○ | ○ | ○ | ○ | ○ |
| » Teaching | ○ | ○ | ○ | ○ | ○ |
| » Scientific writing | ○ | ○ | ○ | ○ | ○ |
| » Giving presentations | ○ | ○ | ○ | ○ | ○ |
| » Writing a grant application | ○ | ○ | ○ | ○ | ○ |
| » Outreach to the public | ○ | ○ | ○ | ○ | ○ |
| » Writing a CV | ○ | ○ | ○ | ○ | ○ |
| » Job applications (academic) | ○ | ○ | ○ | ○ | ○ |
| » Job applications (non-academic) | ○ | ○ | ○ | ○ | ○ |

**GRAD Career Paths**

What stage of grad school are you currently in?

○ Prior qualification exam
○ Post qualification exam

What was your intended career goal when starting grad school?

○ College or university position
○ National/private lab scientist
○ Industry job
○ Teaching K-12
○ [                    ] Other (please describe)
○ Not sure

What is your intended career goal now?

○ » College or university position
○ » National/private lab scientist
○ » Industry job
○ » Teaching K-12
○ [                    ] » Other (please describe)
○ » Not sure

On a scale of 1 to 7, how well do you feel graduate school prepared you for the following career paths?

| | 1 (Not prepared at all) | 2 | 3 | 4 (Moderately prepared) | 5 | 6 | 7 (Extremely prepared) |
|---|---|---|---|---|---|---|---|
| College or university position | ○ | ○ | ○ | ○ | ○ | ○ | ○ |
| National/private lab scientist | ○ | ○ | ○ | ○ | ○ | ○ | ○ |
| Industry job | ○ | ○ | ○ | ○ | ○ | ○ | ○ |
| Teaching k-12 | ○ | ○ | ○ | ○ | ○ | ○ | ○ |
| Other (please describe) [        ] | ○ | ○ | ○ | ○ | ○ | ○ | ○ |

How well were you prepared overall to meet the expectations of your graduate research supervisor(s) when starting your research?

Not well at all
○

Slightly well
○

Moderately well
○

Very well
○

Extremely well
○

When starting grad school, what was your intended area of research?

○ High energy, nuclear, and/or particle experiment
○ High energy, nuclear, and/or particle theory
○ Gravity, astrophysics, and/or cosmology experiment
○ Gravity, astrophysics, and/or cosmology theory
○ [                    ] Other (please describe)

What is/was your eventual type of research in grad school?

○ High energy, nuclear, and/or particle experiment
○ High energy, nuclear, and/or particle theory
○ Gravity, astrophysics, and/or cosmology experiment
○ Gravity, astrophysics, and/or cosmology theory
○ [                    ] Other (please describe)

If you switched research topics from your original expectations and would like to explain why, please feel free to do so.

[                                                                ]

**UNDERGRAD Satisfaction**

How satisfied do you feel with your undergraduate education in physics?

Extremely
dissatisfied
○

Moderately
dissatisfied
○

Slightly
dissatisfied
○

Neither
satisfied
nor
dissatisfied
○

Slightly
satisfied
○

Moderately
satisfied
○

Extremely
satisfied
○

How satisfied do you feel with your undergraduate education in *particle* physics?

Extremely
dissatisfied
○

Moderately
dissatisfied
○

Slightly
dissatisfied
○

Neither
satisfied
nor
dissatisfied
○

Slightly
satisfied
○

Moderately
satisfied
○

Extremely
satisfied
○

**UNDERGRAD Formal Instruction**

When did you first have an opportunity to perform research?

○ Before starting college
○ A first-year research experience (during freshman year)
○ Summer between freshman and sophomore year
○ During sophomore year
○ Summer between sophomore and junior year
○ During junior year
○ Summer between junior and senior year
○ During senior year
○ I have never had a chance to perform any research

How did you first experience your research opportunity?

○ External internship (i.e. SULI, REU, industry-related)
○ With a professor's lab/group at your college
○ Other

How well were you prepared to meet the expectations of your research advisor?

Not well at all
○

Slightly well
○

Moderately well
○

Very well
○

Extremely well
○

What area of particle physics have you had the opportunity to perform research in? (check all that apply)

☐ High energy experiment
☐ High energy theory/phenomenology
☐ Cosmology/particle astrophysics
☐ Not in particle physics (condensed matter, biophysics, computational/mathematical, etc)

Please indicate if you learned about any of following Particle Physics topics during undergrad, inside or outside of the classroom. (check all that apply)

- ☐ Introduction to history and current status of particle physics
- ☐ Introduction to experimental particle physics techniques
- ☐ Introduction to theoretical/phenomenological particle physics techniques
- ☐ Applications to industry and/or engineering
- ☐ Introduction to software/programming used in particle physics (i.e. ROOT, GLoBES, etc...)

For the introductions that you indicated exposure to in the previous question, how did you learn about it? (i.e. format)

| | Classroom | Research | Workshop | Self-guided |
|---|---|---|---|---|
| » Introduction to history and current status of particle physics | ☐ | ☐ | ☐ | ☐ |
| » Introduction to experimental particle physics techniques | ☐ | ☐ | ☐ | ☐ |
| » Introduction to theoretical/phenomenological particle physics techniques | ☐ | ☐ | ☐ | ☐ |
| » Applications to industry and/or engineering | ☐ | ☐ | ☐ | ☐ |
| » Introduction to software/programming used in particle physics (i.e. ROOT, GLoBES, etc...) | ☐ | ☐ | ☐ | ☐ |

Please indicate if you would have liked an introductory experience to each of these topics at your school.

| | Yes | No |
|---|---|---|
| » Introduction to history and current status of particle physics | ○ | ○ |
| » Introduction to experimental particle physics techniques | ○ | ○ |
| » Introduction to theoretical/phenomenological particle physics techniques | ○ | ○ |
| » Applications to industry and/or engineering | ○ | ○ |
| » Introduction to software/programming used in particle physics (i.e. ROOT, GLoBES, etc...) | ○ | ○ |

How well were you prepared overall to meet the mathematical side of your undergraduate physics courses?

Not well at all
○

Slightly well
○

Moderately well
○

Very well
○

Extremely well
○

What undergraduate physics courses did you feel unprepared for mathematically?

☐ Freshmen general physics sequence
☐ Modern Physics/Relativity
☐ Quantum Mechanics
☐ Classical/Theoretical Mechanics
☐ Statistical Mechanics
☐ Electromagnetism
☐ [                    ] Other

What courses outside of physics were you able to take during undergrad which you found useful for your study of physics?

☐ Linear algebra
☐ Group theory
☐ Partial Differential Equations
☐ Programming (C,C++,Python)
☐ Machine learning
☐ Complex Analysis
☐ [                    ] Other

How well were you prepared overall to meet the expectations of your undergraduate physics laboratory courses?

Not well at all
○

Slightly well
○

Moderately well
○

Very well
○

Extremely well
○

**UNDERGRAD Professional Skills**

During undergrad, which of these experiences were you able to participate in? (check all that apply)

☐ Mentor fellow students
☐ Be a part of outreach initiatives
☐ TA work (teaching/grading)
☐ Scientific writing
☐ Presentations/public speaking
☐ CV/resume building

In what setting did your training/experience with each of the following primarily take place?

| | Classroom | Research | Workshop | Self-guided |
|---|---|---|---|---|
| » Mentor fellow students | ☐ | ☐ | ☐ | ☐ |
| » Be a part of outreach initiatives | ☐ | ☐ | ☐ | ☐ |
| » TA work (teaching/grading) | ☐ | ☐ | ☐ | ☐ |
| » Scientific writing | ☐ | ☐ | ☐ | ☐ |
| » Presentations/public speaking | ☐ | ☐ | ☐ | ☐ |
| » CV/resume building | ☐ | ☐ | ☐ | ☐ |

For the experiences you did not participate in, would you have liked to have had the opportunity to do so during undergrad?

|  | Yes | No |
|---|---|---|
| » Mentor fellow students | ○ | ○ |
| » Be a part of outreach initiatives | ○ | ○ |
| » TA work (teaching/grading) | ○ | ○ |
| » Scientific writing | ○ | ○ |
| » Presentations/public speaking | ○ | ○ |
| » CV/resume building | ○ | ○ |

**UNDERGRAD Career Paths**

What year of undergrad are you currently in?

○ 1st year
○ 2nd year
○ 3rd year
○ 4th year
○ 5th year and above

What was your intended career goal when starting undergrad?

○ College or university position
○ National/private lab scientist
○ Industry job
○ Teaching K-12
○ [_____] Other (please describe)
○ Not sure

What is your intended career goal now?

○ College or university position
○ National/private lab scientist
○ Industry job
○ Teaching K-12
○ [_____] Other (please describe)
○ Not sure

Is your next step following graduation finalized? (i.e. if you are in your last year of undergrad, do you know if you have been accepted into a graduate program or have a job waiting?)

○ Yes
○ No
○ Not sure

If your next step is finalized, what is it? If not, what is your intended next step after graduating from undergrad?

○ Graduate school (Physics/Astrophysics)
○ Graduate School (non-Physics/Astro PhD program)
○ Industry job
○ Teaching K-12
○ Professional Field (i.e. Law, Business, Medicine)
○ [                    ] Other (please describe)
○ Not sure

On a scale of 1 to 7, how well do you feel undergraduate school prepared you for the following career paths?

| | 1 (Not prepared at all) | 2 | 3 | 4 (Moderately prepared) | 5 | 6 | 6 (Extremely prepared) |
|---|---|---|---|---|---|---|---|
| College or university position | ○ | ○ | ○ | ○ | ○ | ○ | ○ |
| National/private lab scientist | ○ | ○ | ○ | ○ | ○ | ○ | ○ |
| Industry job | ○ | ○ | ○ | ○ | ○ | ○ | ○ |
| Teaching k-12 | ○ | ○ | ○ | ○ | ○ | ○ | ○ |
| Other | ○ | ○ | ○ | ○ | ○ | ○ | ○ |

**Submit Button**

Thank you for completing the survey! Please press the button to submit.

\clearpage

\end{document}